\documentclass[prb,aps,twocolumn]{revtex4}
\usepackage{bibunits,epsfig,latexsym,amssymb,amsmath,amsbsy,graphics,graphicx,mathrsfs}

\begin{document}
\def \cation{$\kappa$-(ET)$_2$}~
\def \kpx{$\kappa$-(ET)$_2X$}
\def \kbr{$\kappa$-(ET)$_2$Cu[N(CN)$_2$]Br}
\def \deut8br{$\kappa$(d8)-(ET)$_2$Cu[N(CN)$_2$]Br}
\def \h8br{h8-(ET)$_2$Cu[N(CN)$_2$]Br}
\def \kcl{$\kappa$-(ET)$_2$Cu[N(CN)$_2$]Cl}
\def \kncs{$\kappa$-(ET)$_2$Cu(NCS)$_2$}
\def \kcn3{$\kappa$-(ET)$_2$Cu$_2$(CN)$_3$}
\def \>{\textgreater}
\def \<{\textless}
\def \q{\vec{q}}
\def \Q{\vec{Q}}
\def \kpcl{$\kappa$-Cl}
\def \kpbr{$\kappa$-Br}
\def \kpncs{$\kappa$-NCS}
\def \kpcn3{$\kappa$-(CN)$_3$}
\def \d8pbr{$\kappa$(d8)-Br}
\def \m{\mathrm{m}}
\def \max{\mathrm{max}}
\def \cross{\mathrm{cross}}
\def \M{\mathrm{M}}
\def \c{\mathrm{c}}
\def \lw{\mathrm{LW}}
\def \af{\mathrm{AF}}
\def \fm{\mathrm{FM}}
\def \sf{\mathrm{SF}}
\def \res{{\rho \sim T^2}}
\def \us{{\Delta v/v}}
\def \nmr{\mathrm{NMR}}
\def \ks{{K_s}}

\title{Antiferromagnetic Spin Fluctuations and the Pseudogap in the
Paramagnetic Phases of Quasi-Two-Dimensional Organic Superconductors}
\author{Eddy Yusuf, B. J. Powell, and Ross H. McKenzie}
\affiliation{Department of Physics, University of Queensland, Brisbane,
Queensland 4072, Australia}
\date{\today}
\begin{abstract}

The metallic states of a broad range of strongly correlated electron
materials exhibit the subtle interplay between antiferromagnetic spin
fluctuations, a pseudogap in the excitation spectra, and non-Fermi
liquid properties. In order to understand these issues better in the
\cation X family of organic charge transfer salts we give a
quantitative analysis of the published results of nuclear magnetic
resonance (NMR) experiments. The temperature dependence of the nuclear
spin relaxation rate $1/T_1$, the Knight shift $K_s$, and the Korringa
ratio ${\cal K}$, are compared to the predictions of the
phenomenological spin fluctuation model of Moriya, and Millis, Monien
and Pines (M-MMP), that has been used extensively to quantify
antiferromagnetic spin fluctuations in the cuprates. For temperatures
above $T_\nmr \simeq 50$ K, the model gives a good quantitative
description of the data for the paramagnetic metallic phase of several
\cation X materials, with an antiferromagnetic correlation length which
increases with decreasing temperature; growing to several lattice
constants by $T_\nmr$. It is shown that the fact that the dimensionless
Korringa ratio is much larger than unity is inconsistent with a broad
class of theoretical models (such as dynamical mean-field theory) which
neglect spatial correlations and/or vertex corrections. For materials
close to the Mott insulating phase the nuclear spin relaxation rate,
the Knight shift and the Korringa ratio all decrease significantly with
decreasing temperature below $T_\nmr$. This cannot be described by the
M-MMP model and the most natural explanation is that a pseudogap opens
up in the density of states below $T_\nmr$, as in, for example, the
underdoped cuprate superconductors. An analysis of the Mott insulating
phase of \kcn3 is somewhat more ambiguous; nevertheless it suggests
that the antiferromagnetic correlation length is less than a lattice
constant, consistent with a large frustration of antiferromagnetic
interactions as is believed to occur in this material. We show that the
NMR measurements reported for \kcn3~are qualitatively inconsistent with
this material having a ground state with long range magnetic order. A
pseudogap in the metallic state of organic superconductors is an
important prediction of the resonating valence bond theory of
superconductivity. Understanding the nature, origin, and momentum
dependence of the pseudogap and its relationship to superconductivity
are important outstanding problems. We propose specific new experiments
on organic superconductors to elucidate these issues. Specifically,
measurements should be performed to see if high magnetic fields or high
pressures can be used to close the pseudogap.

\end{abstract}

\maketitle

\section{Introduction}

 In the past twenty years a diverse range of new
 strongly correlated electron materials with exotic electronic and
magnetic properties have been synthesized. Examples include
 high-temperature cuprate superconductors,\cite{lee}
manganites with colossal magnetoresistance,\cite{dagotto:cmr}
 cerium oxide catalysts,\cite{esch}
sodium cobaltates,\cite{takada} ruthenates,\cite{
Meano&Mackenzie,grigera} heavy fermion materials,\cite{stewart} and
superconducting organic charge transfer salts.\cite{powell:review} Many
of these materials exhibit a subtle competition between diverse phases:
paramagnetic, superconducting, insulating, and the different types of
order associated with charge, spin, orbital, and lattice degrees of
freedom. These different phases can be explored by varying experimental
control parameters such as temperature, pressure, magnetic field, and
chemical composition. Although chemically and structurally diverse the
properties of these materials are determined by some common features;
such as, strong interactions between the electrons, reduced
dimensionality associated with a layered crystal structure, large
quantum fluctuations, and competing interactions. Many of these
materials are characterized by large antiferromagnetic spin
fluctuations.
 Nuclear magnetic resonance spectroscopy has proven to be a
 powerful probe of local spin dynamics in many
strongly correlated electron
materials.\cite{pennington,nmrnature,sidorov,miyagawa:chemrev} Longer
range and faster spin dynamics have been studied with inelastic neutron
scattering. One poorly understood property of the paramagnetic phases
of many of these materials is the {\it pseudogap} present in large
regions of the phase diagram. Although the pseudogap has received the
most attention in the cuprates,\cite{norman} it is also present in
quasi one-dimensional charge-density wave compounds,\cite{mckenzie:cdw}
manganites,\cite{cmrpseudogap} heavy fermion materials,\cite{sidorov},
in simple metals with no signs of superconductivity,\cite{hayden} and
quite possibly in organic charge transfer
materials.\cite{miyagawa:chemrev} The focus of this paper is on
understanding what information about antiferromagnetic spin
fluctuations and the pseudogap can be extracted from NMR experiments on
the organic charge transfer salts.

The systems which are the subject of the current study are the organic
charge transfer salts based on electron donor molecules BEDT-TTF (ET),
in particular the family \cation X (where $\kappa$ indicates a
particular polymorph\cite{ishiguro}). Remarkably similar physics occurs
in the other dimerised polymorphs, such as the $\beta$, $\beta'$,
$\lambda$, and $\kappa$ phases.\cite{powell:review} These materials
display a wide variety of unconventional behaviours\cite{powell:review}
including: antiferromagnetic and spin liquid insulating states,
unconventional superconductivity, and the paramagnetic metallic phases
which we focus on in this paper. They also share highly anisotropic
crystal and band structures. However, as for various sociological and
historical reasons, the $\kappa$ salts have been far more extensively
studied, and because we intend, in this paper, to make detailed
comparisons with experimental data, we limit our study to $\kappa$
phase salts. This of course begs the question do similar phenomena to
those described below occur in the $\beta$, $\beta'$, or $\lambda$
salts? We would suggest that the answer is probably {\emph yes} but
this remains an inviting experimental question.

\begin{figure*}
\epsfig{file=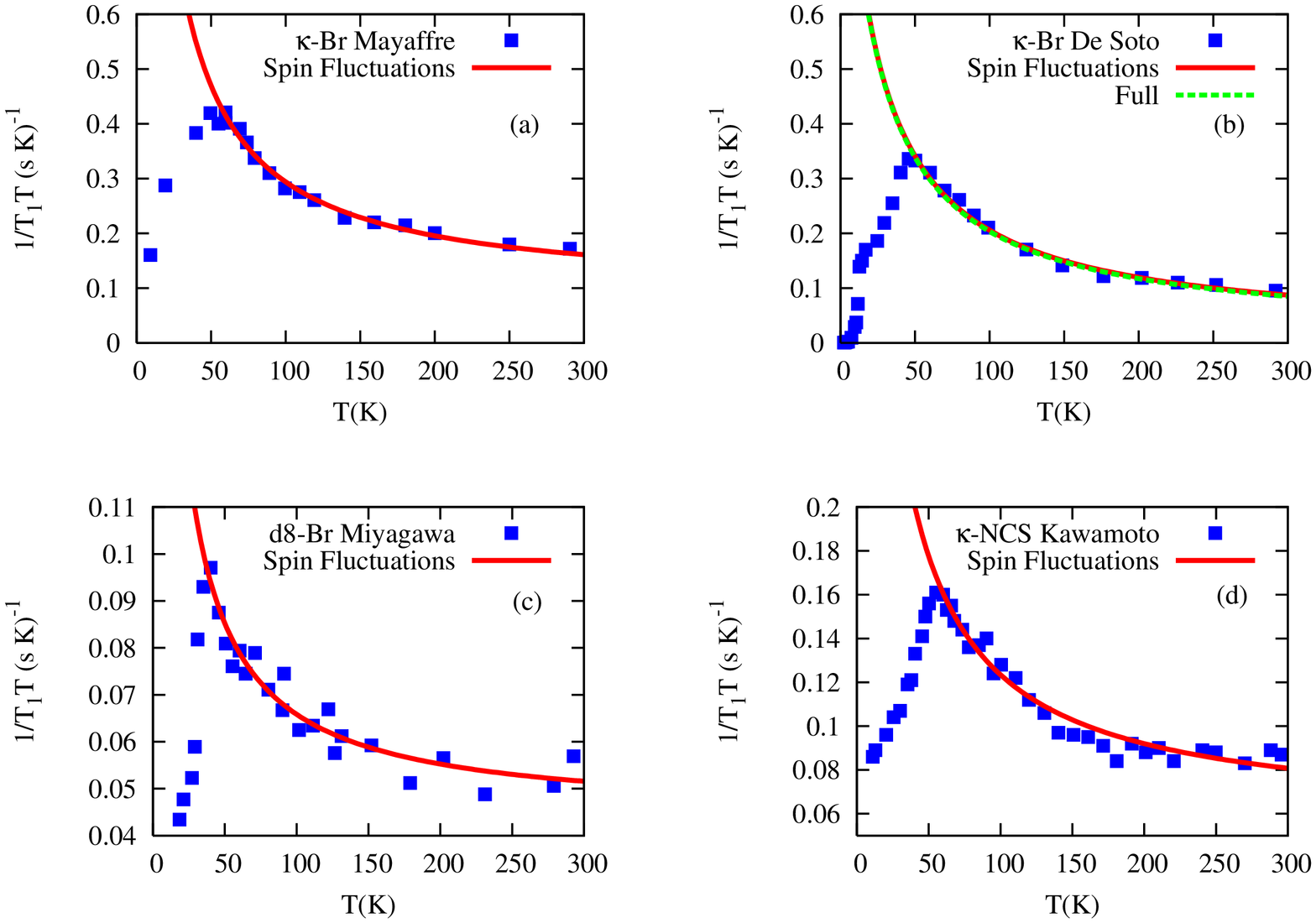,angle=-90,scale=0.6} \caption{[Color
online] Comparison of the measured nuclear spin relaxation rate per
unit temperature, $1/T_1T$, with the predictions of the spin
fluctuation model for various organic charge transfer salts. Panel (a)
shows data for \kbr~measured by Mayaffre {\it et al.}\cite{mayaffre},
panel (b) shows data for \kbr~measured by De Soto {\it et
al.}\cite{desoto}, panel  (c) shows data for \deut8br measured by
Miyagawa {\it et al.}\cite{miyagawa}, and panel (d) shows data for a
\kncs~powder sample measured by Kawamoto {\it et al.}\cite{kawamoto}
The $1/T_1T$ data are weakly temperature dependent at high
temperatures, have a maximum at $T_\nmr \sim 50$ K, and drop abruptly
below $T_\nmr$, contrary to what one would expect for a Fermi liquid in
which $1/T_1T$ is constant. The remarkable similarities of these data
results from the quantitative and qualitative similarity of the
antiferromagnetic spin fluctuations in the paramagnetic metallic phases
of these materials. The parameters that produce the best fits (solid
lines) to Eq. (\ref{nmr_af2}) are tabulated in Table
\ref{tab:parameter}. The spin fluctuation model gives a good fit to the
experimental data between $T_\nmr\sim50$ and room temperature which
suggests strong spin fluctuations in the paramagnetic metallic states
of \kbr, \deut8br, and \kncs. However, below $T_\nmr$ the spin
fluctuation model does not describe the data well. In section
\ref{phase} we argue that this is because a pseudogap opens at
$T_\nmr$. In each figure the solid line indicates the large $T$
approximation of the spin fluctuation model [Eq. (\ref{limiting_t1t})].
To check that this approximation is reasonable we also plot the full
spin fluctuation model [Eq. (\ref{nmr_af2})] as a dashed line in panel
(b). The full and dashed lines cannot be distinguished until well below
$T_\nmr$ and so we concluded that the high $T$ approximation is
excellent in the relevant regime. Note that the analysis on $1/T_1T$
cannot differentiate between antiferromagnetic and ferromagnetic spin
fluctuations (see section \ref{sect:ferro}), but the Korringa ratio
strongly differentiates between these two case and indicates that the
fluctuations are antiferromagnetic (see Fig. \ref{fig:korringa}).  The
nomenclature \kpbr, d8-Br, and \kpncs~is used as shorthand for \kbr,
\deut8br, and \kncs~respectively in the figure keys.}
\label{fig:t1t_fit2}
\end{figure*}

As well as their interesting phenomenology the organic charge transfer
salts are important model systems and a deeper understanding of these
materials may help address a number of important fundamental questions
concerning strongly correlated systems. The (non-interacting) band
structure of the $\kappa$ phase salts is well approximated by the
half-filled tight binding model on the anisotropic
lattice.\cite{powell:review} This model has two parameters $t$, the
hopping integral between nearest neighbor (ET)$_2$ dimers, and $t'$,
the hopping integral between next nearest neighbors across one diagonal
only. In order to describe strongly correlated phases of these
materials we must also include the effects of the Coulomb interaction
between electrons. The simplest model which can include these strong
correlations in the Hubbard model contains one additional parameter
over the tight binding model: $U$, the Coulomb repulsion between two
conduction electrons on the same dimer. In the Hubbard model picture
the different \kpx~salts and different pressures correspond to
different values of $t'/t$ and $U/t$, but all of the $\kappa$ phase
salts are half filled. Varying $U/t$ allows us to tune the proximity to
the Mott transition - understanding the Mott transition and its
associated phenomena remains one most important problems in theoretical
physics\cite{gebhard,imada:rmp} and the organics have provided a new
window on this
problem.\cite{kagawa,merino,limelette,powell:review,kanoda-thermo}
Varying $t'/t$ allows one to tune the degree of frustration in the
system. Understanding the effect of frustration in strongly correlated
systems is of general importance.\cite{takada,Greedan,ong:frustration}
For example there are strong analogies between the organics and
Na$_x$CoO$_2$,\cite{nacoo} and much recent interest has been sparked by
the observation of possible spin liquid states in organic charge
transfer salts.\cite{shimizu,kato,kato2}

\begin{figure}
\begin{center}
\epsfig{file=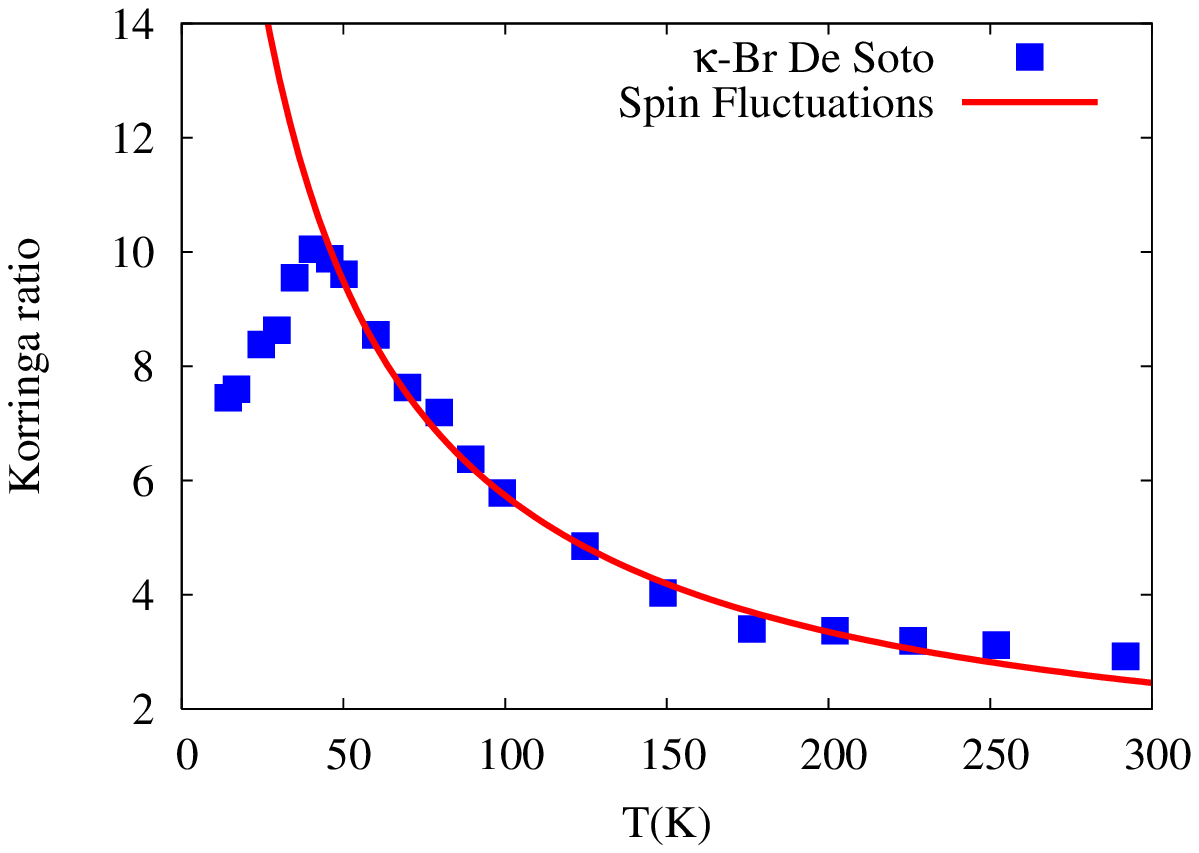,scale=0.7}
\end{center}
\caption{[Color online] Comparison of the Korringa ratio ${\cal
K}\propto1/T_1TK_s^2$ of \kbr~measured by De Soto {\it et
al}.\cite{desoto} with the prediction of the antiferromagnetic spin
fluctuation model. The best fit to Eq. (\ref{korringa-sf}) is indicated
by the solid line. The Korringa ratio is larger than 1 which indicates
that the spin fluctuations are antiferromagnetic (${\cal K}<1$ for
ferromagnetic fluctuations, see Section \ref{sect:ferro}). The
antiferromagnetic correlation length is found to be $3.5\pm2.5$ lattice
spacings at 50 K. Below 50 K the Korringa ratio is suppressed, and in
section \ref{phase} we argue that this is because a pseudogap opens at
$T_\nmr$.  In the key to this figure \kbr~is abbreviated as
\kpbr.}\label{fig:korringa}
\end{figure}

The paramagnetic metallic phases of \kpx~are very different from a
conventional metallic phase. Many features of the paramagnetic metallic
phases agree well with the predictions of dynamical mean field theory
(DMFT) which describes crossover from `bad metal' at high temperatures
to a Fermi liquid as the temperature is
lowered.\cite{merino,Jaime-HH-DMFT,hassan,limelette} This crossover
from incoherent to coherent intralayer\footnote{Throughout this paper
when we discuss coherent versus incoherent behavior we are discussing
the behavior in the planes unless otherwise stated. The subject of the
coherence of transport perpendicular to the layers is a fascinating
issue. We refer the interested reader to one of the reviews on the
subject such as Refs. \onlinecite{Singleton-intra} and
\onlinecite{kartsovnik}.} transport has been observed in a number of
experiments such as resistivity,\cite{limelette}
thermopower,\cite{yu,merino} and ultrasonic
attenuation.\cite{frikach,fournier} The existence of coherent
quasiparticles is also apparent from the observed magnetic quantum
oscillations at low temperatures in
\kpx.\cite{singleton,wosnitza,kartsovnik} However, nuclear magnetic
resonance experiments (reproduced in Figs. \ref{fig:t1t_fit2} and
\ref{fig:korringa}) on the paramagnetic metallic phases on \kpx~do not
find the well known properties of a Fermi liquid. The nuclear spin
relaxation rate per unit temperature, $1/T_1T$, is larger than the
Korringa form predicted from Fermi liquid theory. As the temperature is
lowered $1/T_1T$ reaches a maximum; we label this temperature $T_\nmr$
(the exact value of $T_\nmr$ varies with the anion $X$, but
typically,~$T_\nmr \sim 50$ K, see Fig. \ref{fig:phase}). $1/T_1T$
decreases rapidly as the temperature is lowered below $T_\nmr$ [see Fig
\ref{fig:t1t_fit2}].\cite{mayaffre,desoto,miyagawa:chemrev} The Knight
shift also drops rapidly around $T_\nmr$ [see Fig
\ref{fig:correction}].\cite{desoto} This is clearly in contrast the
Korringa-like behavior one would expect for a Fermi liquid in which
$1/T_1T$ and $K_s$ are constant for $T \ll T_F$, the Fermi temperature.
Similar non-Fermi liquid temperature dependences of $1/T_1T$ and $K_s$
are observed in the cuprates.\cite{timusk,norman:adv} For the cuprates,
it has been argued that the large enhancement of $1/T_1T$ is associated
with the growth of antiferromagnetic spin fluctuation within the
CuO$_2$ planes as the temperature is lowered.\cite{moriya,millis} The
large decrease observed in $1/T_1T$ and $K_s$ is suggestive of a
depletion of the density of states (DOS) at the Fermi level which might
be expected if a pseudogap opens at $T_\nmr$.

A qualitative description of spin fluctuations in the paramagnetic
metallic phase of \kpx~has not been performed previously. The
importance of spin fluctuations in \kpx~below $T_\c$ has been pointed
out by several
groups.\cite{powell:review,schmalian,kino,jujo,ben:prl,kuroki} Since
the nature of the paramagnetic metallic and superconducting phases are
intimately intertwined (in most theories, including BCS,
superconductivity arises from an instability of the metallic phase), it
is important to understand whether the spin fluctuations may extend
beyond the superconducting region into the paramagnetic metallic phase
and if so how strong they are. Another unresolved puzzle is whether
there is a pseudogap in the paramagnetic metallic phase of \kpx~which
suppresses the density of states (DOS) at the Fermi level. A pseudogap
has been suggested on the basis of the NMR data and heat capacity
measurement.\cite{kanoda:jspj2006} If there is a pseudogap then
important questions to answer include: (i) how similar is the pseudogap
in \kpx~to the pseudogaps in the cuprates, manganites and heavy
fermions? (ii) is the pseudogap in \kpx~related to superconductivity?
and (iii) if so how?

\begin{figure*}
\epsfig{file=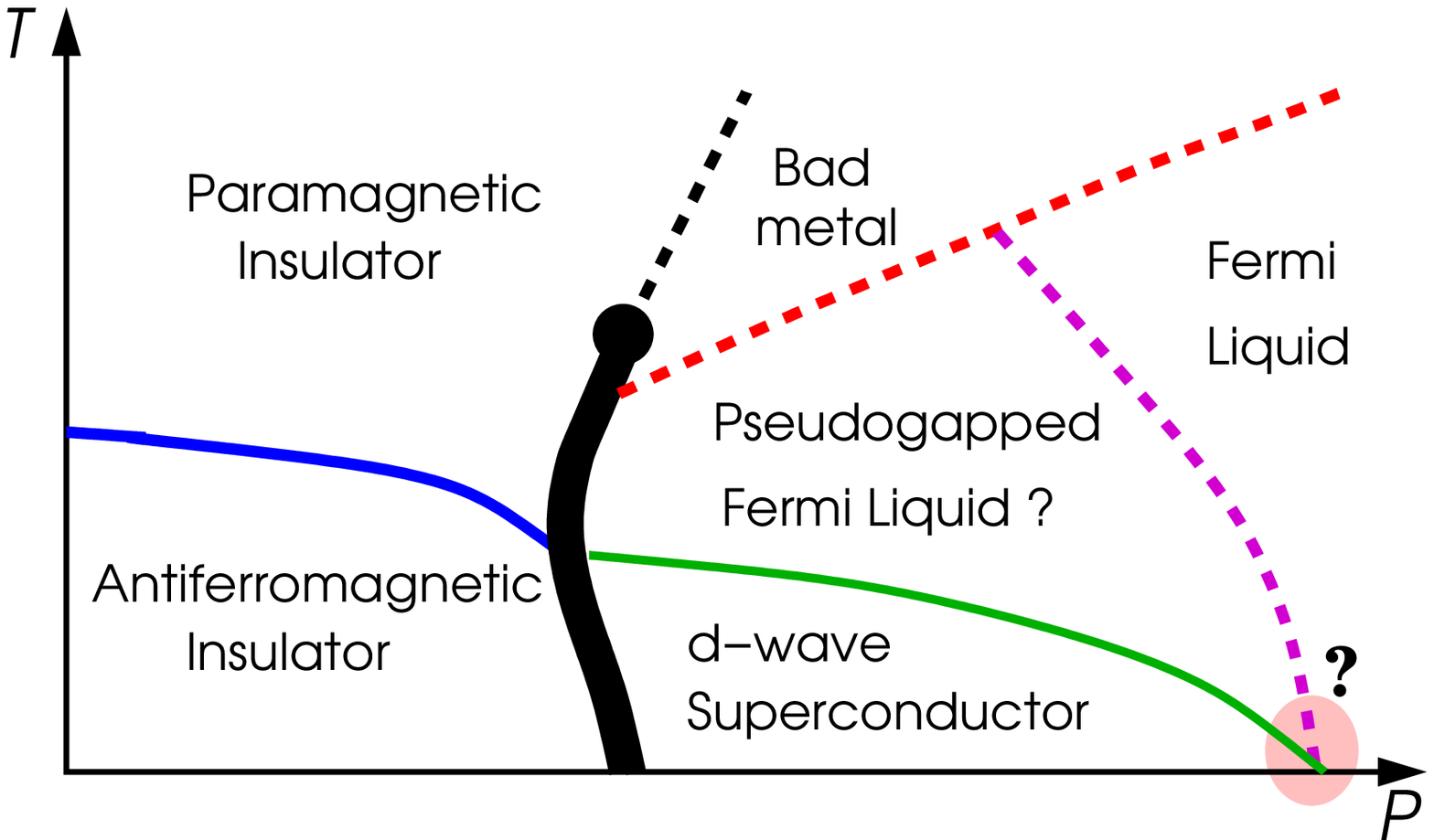,scale=0.7} \caption{[Color online] The
schematic phase diagram for \cation X as a function of temperature and
pressure. Thin solid lines represent second order phase transitions,
the thick solid line is the first order transition line which ends at a
critical point shown as a filled circle, and dashed lines indicate
crossovers. The pseudogap phase is much more complicated than a
renormalized Fermi liquid that has been previously thought to
characterize the paramagnetic metallic phase at low temperatures: it
shows a coherent transport character with long lived quasiparticles,
marked by $T^2$ resistivity behavior\cite{limelette} and magnetic
quantum oscillations,\cite{singleton} but at the same time it exhibits
a loss of density of states  which is clearly seen in the NMR data.
There are not sufficient data at this moment to determine where the
pseudogap phase boundary ends so this uncertainty is represented by the
shaded area with the question mark. More detailed experimental and
theoretical studies in the vicinity of this point will give important
insight into how the pseudogap is related to superconductivity. The
possibility of a quantum critical point somewhere in the vicinity of
the point where the superconducting critical temperature goes to zero
may have important consequences for the observation that the materials
with the lowest superconducting critical temperatures have extremely
small superfluid stiffnesses and are very different from BCS
superconductors.\cite{ujjual,pratt,constraints}}\label{fig:new_phase}
\end{figure*}

A number of scenarios in which a pseudogap may arise have been
proposed. One possible origin of a pseudogap, which a number of
authors\cite{ben:prl, trivedi, zhang,ben:d+id,powell:review} have
argued may be relevant to the organic charge transfer salts, is the
resonating valence bond (RVB) picture (for a review see Refs.
\onlinecite{lee} and \onlinecite{anderson}). In this picture the
electron spins form a linear superposition of spin singlet pairs. The
singlet formation can naturally explain the appearance of a gap: a
non-zero amount of energy is required to break the singlet pairs. For
weakly frustrated lattices, such as the anisotropic triangular lattice
in the parameter range appropriate for \kbr, \kncs, and \kcl, the RVB
theory predicts `$d$-wave'
superconductivity.\cite{ben:prl,kuroki,trivedi,zhang,ben:d+id,powell:review}
This is the symmetry most consistent with a range of experiments on
these
\kpx~salts.\cite{miyagawa:chemrev,powell:prb2004,analytis,powell:group}
However, as the frustration is increased changes in the nature of the
spin fluctuations drive changes in the symmetry of the superconducting
state.\cite{ben:d+id,powell:group} For example, for the isotropic
triangular lattice $t=t'$ [$t \sim t'$ is thought to be appropriate for
\kcn3] RVB theory predicts that the superconducting order parameter has
`$d+id$' symmetry.\cite{ben:d+id,powell:group}

RVB theory also predicts that a pseudogap with the same symmetry as the
superconducting state exists in the paramagnetic phase at temperatures
above the superconducting critical temperature, $T_c$. This pseudogap
results from the formation of short range singlets above $T_c$; only at
$T_c$ do these singlets acquire off-diagonal long-range order. There
are two energy scales ($\Delta$ and $\widetilde \Delta$ in the notation
of, e.g., Refs. \onlinecite{ben:prl} and \onlinecite{zhang}) in the RVB
theory. In the simplest reading of the
theory,\cite{zhang-rmft,anderson} the ratio $\Delta/\widetilde \Delta
\approx T^*/T_c$ where $T^*$ is the temperature at which the pseudogap
opens. For the appropriate model Hamiltonian for the layered \kpx~salts
$\Delta/\widetilde \Delta\simeq5$ near the Mott
transition\cite{ben:prl} which is remarkably similar to the ratio
$T_\nmr/T_c$ in \kbr.

We use the phenomenological antiferromagnetic spin fluctuation model
which was first introduced by Moriya\cite{moriya} and then applied by
Millis, Monien and Pines (MMP)\cite{millis} to cuprates, to examine the
role of spin fluctuations in the paramagnetic metallic phase of \kpx.
We investigate whether the anomalous temperature dependences of NMR
data can be explained without invoking a pseudogap in the DOS. We fit
the spin fluctuation model to the nuclear spin relaxation rate per unit
temperature $1/T_1T$, Knight shift $K_s$, and Korringa ratio
$\mathcal{K}$ data and find that the large enhancements measured in
$1/T_1T$ and $\mathcal{K}$ above $T_\nmr$ are the result of large
antiferromagnetic spin fluctuations [see Figs. \ref{fig:t1t_fit2} and
\ref{fig:korringa}]. The correlation of the antiferromagnetic spin
fluctuations increases as temperature decreases and the relevant
correlation length is found to be $3.5\pm2.5$ lattice spacings at
$T=50$ K. The model produces reasonable agreement with experimental
data down to $T\sim 50$ K. Below 50 K, $1/T_1T$ is suppressed but never
saturates to a constant expected for a Fermi liquid while the spin
fluctuation model produces a monotonically increasing $1/T_1T$ with
decreasing temperature. We argue that this discrepancy between theory
and experiment is due to the appearance of a pseudogap, not captured by
the spin fluctuation model, which suppresses the DOS at the Fermi
level.

Our results suggest the paramagnetic metallic phase of \kpx~is richer
than a renormalized Fermi liquid as has been previously thought to
describe the low temperature metallic state. An exotic regime similar
to the pseudogap in the cuprates appears to be realized in the
paramagnetic metallic phase of \kpx. Thus we believe the appropriate
phase diagram of \kpx~looks like the one sketched in Fig.
\ref{fig:new_phase}. However, the pseudogap in \kpx~is rather peculiar.
On one hand, it shows a coherent intralayer transport (apparent from
the $T^2$ resistivity  and the observed  quantum oscillations). On the
other hand it also shows a loss of DOS apparent from $1/T_1T$ and
$K_s$. Therefore it is important to emphasize that the pseudogap phase
proposed for \kpx~is different from the pseudogap phase realized in the
cuprates. Understanding these differences may well provide important
insight into the physics of the pseudogaps in both classes of material.
We will discuss this matter further in Section V.

We have also applied the spin fluctuation formalism to the Mott
insulating phase \kcn3~(which may have a spin liquid ground
state).\cite{shimizu} While a reasonable agreement between the
calculated and the experimental data on $1/T_1T$ has been obtained, the
result for the Knight shift does not agree with the data. We believe
that this discrepancy reflects the failure of the assumption that the
peak in the dynamic susceptibility dominates even the long wavelength
physics implicit in the spin-fluctuation model. The failure of this
assumption is consistent with the strong frustration in \kcn3~and the
observed spin-liquid behavior of this
material.\cite{shimizu,powell:review} We show that there are
qualitative as well as quantitative difference between the spin
fluctuations in \kcn3~and the spin fluctuations in the other
$\kappa$-phase salts discussed in this paper.

The structure of the paper is as follows. In Section II we introduce
the temperature dependence of the nuclear spin relaxation rate, spin
echo decay rate, Knight shift, and Korringa ratio and describe how they
probe spin fluctuations by probing the dynamical susceptibility. We
calculate these properties in a number of approximations and contrast
the results. In Section III we demonstrate that the spin fluctuation
model provides reasonable fits to the existing experimental results for
\kpx~and discuss its limitations when applied to those materials.
Section IV deals with the application of the spin fluctuation model to
the spin liquid compound \kcn3. We discuss the nature and extent of the
pseudogap in Section V. Finally, in Section VI we suggest new
experiments and give our conclusions.

\section{The Spin Lattice Relaxation Rate,
Spin Echo Decay Rate, and Knight Shift}

The temperature dependence of the nuclear spin lattice relaxation rate
$1/T_1T$, spin echo decay rate $1/T_2$, Knight shift $K_s$, Korringa
ratio ${\cal K}$, and the real and imaginary part of the dynamic
susceptibility, $\chi'({\bf q},\omega)$ and $\chi''({\bf q},\omega)$,
defined by
\begin{equation}
\chi({\bf q},\omega)=\int d^3r~dt e^{i({\bf q} \cdot {\bf r} - \omega
t)} \chi({\bf r},t).
\end{equation}
are discussed in this section. The general expressions for these
quantities are\cite{barzykin}
\begin{subequations}
\label{nmr}
\begin{eqnarray}
\frac{1}{T_1} &=& \lim_{\omega \to 0}\frac{2k_BT}{\gamma_e^2\hbar^4}
\sum_{\bf q} |A({\bf q})|^2\frac{\chi''({\bf q},\omega)}{\omega},\label{t1t}\\
\frac{1}{T^2_2} &=& \frac{2f_A}{\hbar^6 \gamma_e^4}
\sum_{\bf q}|A({\bf q})|^4\chi'({\bf q},0)^2,\\
K_s &=& \frac{|A({\bf 0})| \chi'({\bf0},0)}{\gamma_e\gamma_N
\hbar^2}, \label{ks}\\ {\textrm{and}} \hspace{0.8cm} && \nonumber
\\\mathcal{K} &=& \frac{\hbar}{4\pi
k_B}\left(\frac{\gamma_e}{\gamma_N}\right)^2 \frac{1}{T_1T K_s^2},
\label{korringa}
\end{eqnarray}
\end{subequations}
where $A({\bf q})$ is the hyperfine coupling between the nuclear and
electron spins, $\gamma_N$ ($\gamma_e$) is the nuclear (electronic)
gyromagnetic ratio, and $f_A$ is the relative abundance of the nuclear
spin. For simplicity we will often consider a momentum independent
hyperfine coupling $|A|$ in what follows. Note that Eqs. \ref{nmr} show
that this is an uncontrolled approximation for both $T_1$ and $T_2$,
but that it is not an approximation at all for $K_s$. This is because
$K_s$ only probes the long wavelength physics and hence only depends on
$A({\bf 0})$,  the hyperfine coupling at ${\bf q}=\bf0$.

The calculation of the quantities in Eqs. (\ref{nmr}) boils down to
determining the appropriate form of the dynamic susceptibility. In
the following sections we  begin by discussing the role of vertex
corrections in determining the properties measured by NMR (section
\ref{sect:vertex}), before moving onto the a variety of
approximations for calculating the dynamic susceptibility. They are
antiferromagnetic and ferromagnetic spin fluctuations (section
\ref{sect:sf}), dynamical mean field theory (section
\ref{sect:dmft}), and the $1/N$ approach to the quantum critical
region of frustrated two-dimensional antiferromagnets (section
\ref{sect:qcp}). However, in this work we will predominantly use the
spin fluctuation model to analyze the NMR data. The other models are
presented for comparison.

\subsection{The Spin Fluctuation Model}\label{sect:sf}

The dynamic susceptibility in this model is written
as\cite{moriya,millis}
\begin{equation}
\chi({\bf q},\omega) = \chi_\lw(\omega) + \chi_\af({\bf q},\omega),
\label{dynamic}
\end{equation}
where $\chi_\lw(\omega)$ is the dynamic susceptibility in the long
wavelength regime and $\chi_\af({\bf q},\omega)$ is a contribution to
the dynamic susceptibility which peaks at some wave vector ${\bf Q}$.
These susceptibilities take the form
\begin{eqnarray}
\chi_\lw(\omega) &=& \frac{\bar{\chi}_0(T)}{1-i\omega/\Gamma(T)}\nonumber\\
\chi_\af({\bf q},\omega) &=& \frac{\chi_Q(T)}{1+\xi(T)^2|{\bf
q}-{\bf Q}|^2-i\omega/\omega_\sf(T)} \label{dynamic_af}
\end{eqnarray}
where $\bar{\chi}_0(T)$ [$\chi_Q(T)$] is the static spin susceptibility
at ${\bf q}={\bf 0}$ [${\bf Q}$], $\Gamma(T)$ [$\omega_\sf(T)$] is the
characteristic spin fluctuation energy which represents damping in the
system near ${\bf q}={\bf0}$ [${\bf Q}$], and $\xi(T)$ is the
temperature dependent correlation length. Hence, the real and imaginary
parts of the dynamic susceptibility can then be written as
\begin{eqnarray}
\chi'({\bf q},0) &=& \bar{\chi}_0(T) \left[1+\frac{\chi_Q(T)}
{\bar{\chi}_0(T)}\frac{1}{(1+\xi(T)^2|{\bf q}-{\bf Q}|^2)^2}\right]\nonumber\\
\chi''({\bf q},\omega) &=& \frac{\omega
\bar{\chi}_0(T)}{\Gamma}\nonumber\\
&& \left[1+\frac{\chi_Q(T) \Gamma}{\bar{\chi}_0(T) \omega_\sf(T)}
\frac{1}{(1+\xi(T)^2|{\bf q}-{\bf Q}|^2)^2}\right].
\nonumber\\\label{real_im_af}
\end{eqnarray}
Note that the above form of $\chi_{LW}(\omega)$ is the appropriate form
for a Fermi liquid. Therefore if the system under discussion is not a
Fermi liquid then the validity this expression for $\chi_{LW}(\omega)$
cannot be guaranteed. For example, the marginal Fermi liquid theory
predicts a different frequency dependence.\cite{varma} If the dynamic
susceptibility is sufficiently peaked then $1/T_1$ will not be strongly
dependent on the long wavelength physics [because $1/T_1$ measures the
susceptibility over the entire Brillouin zone, c.f., Eq. (\ref{t1t}),
and therefore will be dominated by the physics at ${\bf q}=\bf Q$]. On
the other hand the Knight shift is a measure of the long wavelength
properties [c.f., Eq. (\ref{ks})] and therefore may be sensitive to the
details of $\chi_\lw(\omega)$. Here we will explicitly assume,
following the assumption made by MMP\cite{millis}, that the uniform
susceptibility $\bar{\chi_0}$ and the spin fluctuation energy near
$q=0$ is temperature independent. One justification for this
approximation in organics is that the Knight shift is not strongly
temperature dependent [see Section III D]. This approximation will
break down in the systems where the uniform susceptibility is strongly
temperature dependent such as
YBa$_2$Cu$_3$O$_{6.63}$\cite{monien:prb43} and
La$_{1.8}$Sr$_{0.15}$CuO$_4$.\cite{monien:prb43b}

In the critical region  $\xi(T) \gg a$, where $\xi(T)$ is the
correlation length, and $a$ is the lattice constant, and one
has\cite{millis}
\begin{eqnarray}
\chi_Q(T) &=& \left(\frac{\xi(T)}{\xi_0}\right)^{2-\eta}
\bar{\chi_0}\nonumber\\
\omega_\sf(T) &=& \left(\frac{\xi_0}{\xi(T)}\right)^{z}\Gamma
\end{eqnarray}
where $\eta$ is the critical exponent which governs the power-law decay
of the spin correlation function, $z$ is the dynamical critical
exponent, and $\xi_0$ is some temperature independent length scale. The
simplest possible assumptions are relaxation dynamics for the spin
fluctuations (which are characterized by $z=2$) and mean field scaling
of the spin correlations ($\eta=0$). With these approximations the real
and imaginary parts of the dynamic susceptibility are given by
\begin{eqnarray}
\chi'({\bf q},0)&=&\bar{\chi}_0 \left[1+\sqrt{\beta}\frac{[\xi(T)/a]^2}
{[1+\xi(T)^2|{\bf q}-{\bf Q}|^2]^2}\right]\nonumber\\
 \chi''({\bf q},\omega) &=& \frac{\omega \bar{\chi}_0}{\Gamma}
\left[1+\beta \frac{[\xi(T)/a]^4}{[1+\xi(T)^2|{\bf q}-{\bf
Q}|^2]^2}\right]
\end{eqnarray}
where $\beta=(a/\xi_0)^4$. The temperature independent, dimensionless
parameter $\beta$ can also be expressed in terms of the original
variables appearing in the dynamic susceptibility in Eq.
(\ref{dynamic_af}) as
\begin{equation}
\beta=\frac{\chi_Q(T) \Gamma}{\bar{\chi}_0 \omega_\sf(T)}
\left(\frac{a}{\xi(T)}\right)^4.
\end{equation}
Written in this form, $\beta$ has a clear interpretation: it represents
the strength of the spin fluctuations at a finite wave vector $\bf Q$.
We will now consider two cases: antiferromagnetic and ferromagnetic
spin fluctuations.

\subsubsection{Antiferromagnetic Spin Fluctuations}

If we have antiferromagnetic spin fluctuations then the dynamic
susceptibility $\chi({\bf q},\omega)$ peaks at a finite wave vector
${\bf q}={\bf Q}$; for example, for a square lattice ${\bf
Q}=(\pi,\pi)$. The NMR relaxation rate, spin echo decay rate, and
Knight shift can be calculated straightforwardly from the real and
imaginary parts of the dynamic susceptibility given in Eq.
(\ref{real_im_af}). The results are
\begin{subequations}
\label{nmr_af}
\begin{eqnarray}
\frac{1}{T_1T} &=&\frac{2\pi k_B |A|^2
\bar{\chi}_0}{\gamma_e^2\hbar^4\Gamma}
\left[1+\beta\frac{[\xi(T)/a]^4}
{1+[\tilde{Q}\xi(T)]^2}\right]\label{nmr_af_t1t}\\
\frac{1}{T_2^2} &=& \frac{2f_A |A|^4\bar{\chi}_0^2}{\pi \gamma_e^4
\hbar^6} \bigg[1+\beta\frac{[\xi(T)/a]^4} {1+[\tilde{Q}\xi(T)]^2}\nonumber\\
&&+ \frac{\sqrt{\beta}}{\pi}\ln(1+[\tilde{Q}\xi(T)]^2)\bigg]
\label{nmr_af_t2}
\\
K_s &=& \frac{|A|\bar{\chi}_0}{\gamma_e \gamma_N \hbar^2}\left[1 +
\sqrt{\beta} \frac{[\xi(T)/a]^2}{1+[\tilde{Q}\xi(T)]^2}\right]
\label{nmr_af_ks}\\
{\cal K} &=&
\frac{\hbar\gamma_e^2}{2\Gamma\bar{\chi}_0}\frac{\left[1+\beta\frac{[\xi(T)/a]^4}
{1+[\tilde{Q}\xi(T)]^2}\right]}{\left[1 + \sqrt{\beta}
\frac{[\xi(T)/a]^2}{1+[\tilde{Q}\xi(T)]^2}\right]^2},\label{nmr_af_korringa}
\end{eqnarray}
\end{subequations}
where $\tilde{Q}$ is a cutoff from the momentum integration [c.f. Eq.
(\ref{t1t})]. For $\xi(T) \gg a$: $1/T_1T \sim \xi(T)^2$, $1/T_2 \sim
\xi(T)$, and $K_s \sim$ constant which leads to the Korringa ratio
$\mathcal{K} \simeq (\hbar\gamma_e^2/2\Gamma\bar{\chi}_0)
[\tilde{Q}\xi(T)]^2$. In this model the Korringa ratio can only be
equal to unity if the spin fluctuations are completely suppressed
($\beta=0$). Hence, one expects ${\cal K}>1$ if antiferromagnetic
fluctuations are dominant.\cite{moriya:jpsj,narath} This indicates that
there are significant vertex corrections when there large strong
antiferromagnetic fluctuations.


\subsubsection{Ferromagnetic Spin Fluctuations}\label{sect:ferro}

For ferromagnetic spin fluctuations, $\chi({\bf q},\omega)$ is peaked
at ${\bf q}=\bf0$. The NMR relaxation rate and spin echo decay rate are
exactly the same as those given in Eqs. (\ref{nmr_af_t1t}) and
(\ref{nmr_af_t2}) because $1/T_1T$ and $1/T_2$ come from summing the
contributions form all wave vectors in the first Brillouin zone, which
makes the location of the peak in q space irrelevant. In contrast, the
Knight shift will be different in the ferromagnetic and
antiferromagnetic cases because $K_s$ only measures the ${\bf q}=\bf0$
part of the dynamic susceptibility; it will be enhanced by the
ferromagnetic fluctuations. Thus, in the ferromagnetic spin fluctuation
description $K_s$ is given by
\begin{eqnarray}
K_s &=& \frac{|A|\bar{\chi}_0}{\gamma_e \gamma_N \hbar^2}\left[1 +
\sqrt{\beta} (\xi/a)^2\right]\label{nmr_fm}
\end{eqnarray}
and
\begin{eqnarray}
{\cal K} &=& \frac{\hbar\gamma_e^2}{2\Gamma\bar{\chi}_0}
\frac{\left[1+\beta\frac{[\xi(T)/a]^4}{1+[\tilde{Q}\xi(T)]^2}\right]}{\left[1
+ \sqrt{\beta} (\xi/a)^2\right]^2}\label{nmr_fm_korringa}
\end{eqnarray}
For $\xi(T) \gg a$: $1/T_1T \sim \xi(T)^2$, $1/T_2 \sim \xi(T)$, and
$K_s \sim \xi(T)^2$ which leads to  $\mathcal{K} \simeq
(\hbar\gamma_e^2/2\Gamma\bar{\chi}_0) [\pi\xi(T)/a]^{-2}$. We can see
that $\mathcal{K}<1$ in the presence of ferromagnetic
fluctuations.\cite{moriya:jpsj,narath} So again vertex corrections are
important if the system has strong ferromagnetic fluctuations. Recall
that, in contrast, in the case of antiferromagnetic fluctuations the
Korringa ratio is larger than one. Thus analyzing the Korringa ratio
allows one to straightforwardly distinguish between antiferromagnetic
and ferromagnetic spin fluctuations.

\subsection{Dynamical Mean Field Theory}\label{sect:dmft}

DMFT is an approach based on a mapping of the Hubbard model onto a
self-consistently embedded Anderson impurity
model.\cite{georges,kotliar,pruschke} DMFT predicts that the metallic
phase of the Hubbard model has two regimes with a crossover from one to
the other at a temperature $T_0$. For $T\<T_0$ the system is a
renormalized Fermi liquid characterized by Korringa-like temperature
dependence of $1/T_1T$ and coherent intralayer transport. Above $T_0$,
the system exhibits anomalous properties with $1/T_1T \sim a +
b(T_0/T)$ (c.f., Ref. [\onlinecite{pruschke}]) and incoherent charge
transport. This regime is often refereed to as the `bad metal'.
Microscopically the bad metal is characterized by quasi-localized
electrons and the absence of quasiparticles. Thus DMFT predicts that at
high temperatures $1/T_1T \sim a + b(T_0/T)$, but $1/T_1T$ saturates to
a constant below $T_0$. This temperature dependence is similar to that
for the single impurity Anderson model.\cite{jarrell} Note that this
temperature dependence is qualitatively similar to that found for spin
fluctuations [c.f., Eq. (\ref{limiting_t1t}), the high temperature
limit of Eq. (\ref{nmr_af_t1t})].

The predictions of DMFT correctly describe the properties of a range of
transport and thermodynamic experiments on the organic charge transfer
salts.\cite{merino,limelette,powell:review} This suggests that these
systems undergo a crossover from a bad metal regime for $T \> T_0$ to a
renormalized Fermi liquid below $T_0$. As we shall see in more detail
later (see Fig. \ref{fig:t1t_fit2}) the nuclear spin relaxation rate is
suppressed but never saturates below $T_\nmr$; this  is \emph{not}
captured by DMFT. This suggests that the low-temperature regime of
\kpx~is more complicated than the renormalized Fermi liquid predicted
by DMFT which until now, is widely believed to be the correct
description of the low temperature paramagnetic metallic state in the
organic charge transfer salts. We will discuss the reasons for and
implications of the failure of DMFT to correctly describe NMR
experiments on the organic charge salts in section \ref{phase}.

\subsection{Quantum Critical Region of Frustrated 2D Antiferromagnets}\label{sect:qcp}

The static uniform and dynamic susceptibilities of nearly-critical
frustrated 2D antiferromagnets has been studied by Chubukov {\it et
al.}\cite{chubukov} They considered a long-wavelength action with an
$N$ component, unit-length, complex vector which has a
SU($N$)$\otimes$O(2) symmetry and performed a $1/N$ expansion. This
gives susceptibilities which follow a universal scaling form.
%
As one approaches the quantum critical point, the spin stiffness will
vanish but the ratio between in-plane and out-of-plane stiffnesses
remains finite and approaches unity. 
In this regime, to
order $1/N$, the quantities in Eq. (\ref{nmr}) are given
by\cite{chubukov}
\begin{equation}
1/T_1 \sim T^{\eta};~~ 1/T_2 \sim T^{\eta-1};~~ K_s \sim T;~~
\mathcal{K} \sim T^{\eta-2}.\label{nmr_qc}
\end{equation}

\section{Spin Fluctuations in \kpx}

The NMR relaxation rate per unit temperature, Knight shift, and
Korringa ratio in the antiferromagnetic spin fluctuations model are
given by Eqs. (\ref{nmr_af_t1t}), (\ref{nmr_af_ks}), and
(\ref{nmr_af_korringa}). Their temperature dependence comes through the
antiferromagnetic correlation length. We adopt the form of $\xi(T)$
from M-MMP\cite{moriya,millis}: $\xi(T)/\xi(T_x) =
\sqrt{2T_x/(T+T_x)}$. For this form of the correlation length $T_x$
represents a natural temperature scale and $\xi(T)$ is only weakly
temperature dependent for $T\ll T_x$. For this choice of $\xi(T)/a$ we
have
\begin{eqnarray}
\frac{1}{T_1T} &=& \left(\frac{1}{T_1T}\right)_0 \left[1+\frac{\beta
C^2}{(T/T_x+1)^2+2\pi^2C(T/T_x+1)}\right]\nonumber\\
K_s &=& (K_s)_0\left[1+\frac{\sqrt{\beta} C
}{1+2\pi^2C+T/T_x}\right]\nonumber\\
{\cal K} &=& {\cal K}_0 \frac{\left[1+\frac{\beta
C^2}{(T/T_x+1)^2+2\pi^2C(T/T_x+1)}\right]}{\left[1+\frac{\sqrt{\beta} C
}{1+2\pi^2C+T/T_x}\right]^2},
 \label{nmr_af2}
\end{eqnarray}
where we have defined
\begin{eqnarray}
C&=&2\left[\frac{\xi(T_x)}{a}\right]^2,\nonumber\\
(1/T_1T)_0&=&\frac{2\pi k_B |A|^2 \bar{\chi}_0}{\gamma_e^2\hbar^4\Gamma},\nonumber\\
(K_s)_0&=&\frac{|A|\bar{\chi}_0}{\gamma_e \gamma_N
\hbar^2},\nonumber\\ \textrm{and}~~ {\cal K}_0 &=&
\frac{\hbar\gamma_e^2}{2\Gamma\bar{\chi}_0},\label{coefficient}
\end{eqnarray}
to simplify the notation.

\subsection{The Nuclear Spin Relaxation Rate}

We now analyze the temperature dependence of $1/T_1$. In the discussion
to follow, we will assume that the correlation length is large compared
to unity and to $T/T_x$, i.e, $C =2(\xi(T_x)/a)^2 \gg T/T_x$. By this
assumption, the limiting cases of $1/T_1T$ [Eq. (\ref{nmr_af2})] are
given by
\begin{subequations}
\label{limit_t1t}
\begin{eqnarray}
\frac{(T_1T)_0}{T_1T} &\simeq&
1+\frac{\beta}{\pi^2}\left(\frac{\xi(T_x)}{a}\right)^2\left[1 -
\left(\frac{T}{T_x}\right)\right]\nonumber\\
&&\mathrm{for}~T \ll T_x\\
 \frac{(T_1T)_0}{T_1T} &\simeq& 1 + \frac{\beta}{\pi^2}
 \left(\frac{\xi(T_x)}{a}\right)^2
\left(\frac{1}{T/T_x+1}\right)\nonumber\\
&&\mathrm{for}~T \gg T_x. \label{limiting_t1t}
\end{eqnarray}
\end{subequations}
The NMR relaxation rate per unit temperature calculated from the spin
fluctuation model is a monotonic function of temperature. In the
high-temperature regime $1/T_1T$ has a $T^{-1}$ dependence while in the
low-temperature regime it is linear in $T$ with a \emph{negative}
slope. Thus one realizes immediately that the data for temperatures
below $T_\nmr$ is \emph{not} consistent with the predictions of the
spin fluctuation theory as it has a \emph{positive} slope. We will
return to discuss this regime latter. We begin by investigating the
high temperature regime, $T>T_\nmr$.


We fit the $1/T_1T$ expression for $T \gg T_x$ (\ref{limiting_t1t}) to
the experimental data of De Soto\cite{desoto} for \kbr~ between
$T_\nmr$ and 300 K with $(1/T_1T)_0$, $\beta [\xi(T_x)/a]^2$, and $T_x$
as free parameters. It is not possible to obtain $\beta$ and
$\xi(T_x)/a$ unambiguously from fitting to $1/T_1T$ data because the
model depends sensitively only on the product $\beta [\xi(T_x)/a]^2$
(see Eq. (\ref{limiting_t1t})). The parameters from the fits are
tabulated in Table \ref{tab:parameter} and the results are plotted in
Fig. \ref{fig:t1t_fit2}. The use of Eq. (\ref{limiting_t1t}) to fit
$1/T_1T$ data is justified {\it post hoc} since $T_x$ is found to be
$2-6$ times smaller than $T_\nmr$. We have also checked this by
plotting the full theory (without taking the $T\gg T_x$ limit) for
\kbr~in Fig. \ref{fig:t1t_fit2}b, where there is Korringa ratio data
(see Fig. \ref{fig:korringa}) and thus we can determine $\beta$ and
$\xi(T_x)/a$ individually. It can be seen from Fig. \ref{fig:t1t_fit2}b
that the disagreement between the full theory and the high temperature
approximation is smaller than the thickness of the curves, therefore
this approximation is well justified. It will be shown in Section II B
that the correlation length is indeed large thus providing further
justification for the use of Eq. (\ref{limiting_t1t}) here.

The model produces a reasonably good fit to the experimental data on
\kbr\cite{desoto} between $T_\nmr$, the temperature at which $1/T_1T$
is maximum, and room temperature. In the high temperature regime (e.g.,
around room temperature), $1/T_1T$ has a very weak temperature
dependence, indicating weakly correlated spins. The large enhancement
of $1/T_1T$ can be understood in terms of the growth of the spin
fluctuations: as the system cools down, the spin-spin correlations grow
stronger which allows the nuclear spins to relax faster by transferring
energy to the rest of the spin degrees of freedom via these spin
fluctuations. Strong spin fluctuations, measured by large values of
$\beta [\xi(T_x)/a]^2$, are not only present in \kbr~but also observed
in other materials such as the fully deuterated \kbr~\{which will be
denoted by \deut8br\} and \kncs. The results of the fits for
\deut8br~and \kncs~are shown in Fig. \ref{fig:t1t_fit2}. The parameters
that produce the best fits are also tabulated in Table
\ref{tab:parameter}. This suggests that strong spin fluctuations are
present in these charge transfer salts. In all cases studied here,
strong spin fluctuations are evident from the large value of $\beta
[\xi(T_x)/a]^2$.

The nature of the spin fluctuations, i.e., whether they are
antiferromagnetic or ferromagnetic, cannot, even in principle, be
determined from the analysis on $1/T_1T$. Both cases yield the same
$1/T_1T$ [see Eq. (\ref{nmr_af}) and Sec II.C.2] because the nuclear
spin relaxation rate is obtained by summing all wave vector
contribution in the first Brillouin zone. However, in the next section
we will use the Korringa ratio to show that the spin fluctuations are
antiferromagnetic.

\begin{table}
\begin{tabular}{c | c | c | c | c}
\hline\hline Material & Ref. & $(1/T_1T)_0$ & $T_x$ (K)& $\beta [\xi(T_x)/a]^2$\\
& & ($s^{-1}$K$^{-1}$) &  &\\
\hline
\kpbr & Mayaffre [\onlinecite{mayaffre}] & 0.09 $\pm$ 0.01 & 6.5 $\pm 5.5$ & 290 $\pm 250$\\
\kpbr& De Soto [\onlinecite{desoto}] & 0.02 $\pm$ 0.01 & 20 $\pm 10$ & 680 $\pm 430$\\
d8-Br &Miyagawa [\onlinecite{miyagawa}] & 0.04 $\pm 0.01$ & 6.2 $\pm 3.5$ & 85 $\pm 65$\\
\kpncs& Kawamoto [\onlinecite{kawamoto}] & 0.06 $\pm 0.01$ & 11 $\pm
2.6$ & 110 $\pm 89$\\\hline\hline
\end{tabular}
\caption{The parameters obtained from the fits which are used to
produce Fig. \ref{fig:t1t_fit2}. Evidence for strong spin fluctuations
come from the large value of $\beta [\xi(T_x)/a]^2$ which are present
for all the materials tabulated above. In the table \kpbr, d8-Br, and
\kpncs~are used as shorthand for \kbr, \deut8br, and
\kncs~respectively.} \label{tab:parameter}
\end{table}

Below $T_\nmr$, the calculated $1/T_1T$ continues to rise while the
experimental data shows a decrease in the nuclear spin relaxation rate
per unit temperature. However, the data does not reach a constant
$1/T_1T$ as expected for a Fermi liquid. This indicates that the
physics below $T_\nmr$ is dominated by some other mechanism not
captured by the spin fluctuation, Fermi liquid, or DMFT theories. One
possibility is a pseudogap opens up at $T_\nmr$ which suppresses the
DOS at the Fermi level. Since $1/T_1T \sim \tilde{\rho}^2(E_F)$ [c.f.,
Eq. (\ref{t1t_local2})], a decrease in DOS will naturally lead to the
suppression of $1/T_1T$. One might argue that the discrepancy between
the theory and experiments below $T_\nmr$ stems from our assumption of
a q-independent hyperfine coupling in the $1/T_1T$ expression. However,
in section \ref{knight-shift} we will show that the Knight shift is
also inconsistent with the predictions of the spin fluctuation model
below $T_\nmr$. While including the appropriate q-dependence of the
hyperfine coupling might change the temperature dependence of $1/T_1T$,
it certainly cannot affect the temperature dependence of the Knight
shift as can be seen from Eq. (\ref{ks}).

\subsection{The Korringa Ratio}

In the previous section we compared the prediction of the spin
fluctuation model for $1/T_1T$ to the experimental data and obtained
good agreement with the data between $T_\nmr$ and 300 K. However, we
were not able to determine $\beta$ and $\xi(T_x)/a$ unambiguously
because $1/T_1T$ is sensitive only to the product $\beta
[\xi(T_x)/a]^2$. We were also not able to determine whether
antiferromagnetic or ferromagnetic spin fluctuations are dominant. We
resolve these by studying the Korringa ratio ${\cal K}$. It has
previously been pointed out that antiferromagnetic (ferromagnetic)
fluctuations produce a Korringa ratio that is larger (less) than
one.\cite{moriya:jpsj,narath} We have also shown in Section II that in
the limit of large correlation length, the Korringa ratio behaves like
$(\xi/a)^2 > 1$ for antiferromagnetic spin fluctuations and like
$(a/\xi)^2 < 1$ for ferromagnetic spin fluctuations. The Korringa ratio
data for \kbr~(see Fig \ref{fig:korringa}) is significantly larger than
one at all temperatures which shows that antiferromagnetic fluctuations
dominate. With this in mind, we study the antiferromagnetic spin
fluctuation model.

First we note that $K_s$, given by Eq. (\ref{nmr_af2}), has a weak
temperature dependence because $C=2(\xi(T_x)/a)^2$ is generally larger
than unity and $T/T_x$. In the limit of large correlation lengths, the
second term inside the square bracket in the expression for $K_s$ given
in Eq. (\ref{nmr_af2}) can be approximated as $(1+C+T/T_x)^{-1} \simeq
C^{-1}$ and the Knight shift will be given by $K_s \simeq (K_s)_0
(1+\sqrt{\beta}/(2\pi^2))$ which is temperature independent. We use
this temperature independent Knight shift to calculate the Korringa
ratio $\cal{K}$
\begin{eqnarray}
{\cal K} &=& \frac{\hbar}{4\pi
k_B}\left(\frac{\gamma_e}{\gamma_N}\right)^2\frac{1}{T_1T K_s^2}
\label{korringa-sf}\\\nonumber &\simeq& {\cal
K}_0\left(1+\frac{\beta[\xi(T_x)/a]^2}
{\pi^2(T/T_x+1)}\right)\left(\frac{1}{1+\sqrt{\beta}/(2\pi^2)}\right)^2
\end{eqnarray}
where the prefactor ${\cal K}_0$ is given by Eq.
(\ref{coefficient}).

We fit Eq. (\ref{korringa-sf}) to the experimental Korringa data for
\kbr.\cite{desoto} The result is plotted in Fig. \ref{fig:korringa}. In
this expression we have three parameters, $\beta [\xi(T_x)/a]^2$,
$T_x$, and $\sqrt{\beta}$, two of which, $\beta [\xi(T_x)/a]^2$ and
$T_x$, have been determined from fitting $1/T_1T$. There is only one
remaining free parameter in the model, $\sqrt{\beta}$, which can then
be determined unambiguously from the Korringa fit which yields
$\beta=60 \pm 20$. This value of $\beta$ implies that the
antiferromagnetic correlation length $\xi(T) = 3.5 \pm 2.5 a$ ($a$ is
the unit of one lattice constant) at $T = 50$ K. This value is in the
same order of magnitude as the value of the correlation length
estimated in the cuprates.\cite{monien:prb43}

The Korringa ratio data are well reproduced by the antiferromagnetic
spin fluctuation model when $T>T_\nmr$. This is again consistent with
our earlier conclusion that the spin fluctuations have
antiferromagnetic correlations. A large Korringa
ratio\cite{takigawa:physica_c,bulut} has previously been observed in
the cuprates indicating similar antiferromagnetic fluctuations in these
systems. The Korringa ratio has also been measured in a number of heavy
fermion compounds\cite{ishida,aart,kitaoka} Similar antiferromagnetic
fluctuations, like those observed in the cuprates and organics, are
present in CeCu$_2$Si$_2$ and. The Korringa ratio of this material has
a value of 4.6 at 100 mK (Ref. [\onlinecite{aart}]).
In contrast, YbRh$_2$Si$_2$\cite{ishida} and
CeRu$_2$Si$_2$\cite{kitaoka}, show strong ferromagnetic spin
fluctuations as is evident from the Korringa ratio less than unity. In
Sr$_2$RuO$_4$\cite{ishida:ruthenates} the Korringa ratio is
approximately 1.5 at 1.4 K. Upon doping with Ca to form
Sr$_{2-x}$Ca$_x$$_2$RuO$_4$, the Korringa ratio becomes less than one
which indicates that there is a subtle competition between
antiferromagnetic and ferromagnetic fluctuations in these ruthenates.

\subsection{The Antiferromagnetic Correlation Length}

It is important to realize that the spin fluctuation formalism can be
used to extract quantitative information about the spin correlations
from NMR data. For example, the fits presented in Figs.
\ref{fig:t1t_fit2} and \ref{fig:korringa} allow us to estimate the
antiferromagnetic correlation length. From the fit for \kbr~(Table
\ref{tab:parameter}) we found that the antiferromagnetic correlation
length $\xi(T)/a= 3.5 \pm 2.5$ at $T=T_\nmr=50$~K. In order to
understand the physical significance of this value of $\xi(T)$ it is
informative to compare this value with the correlation length for the
square\cite{ding} and triangular\cite{elstner} lattice
antiferromagnetic Heisenberg models.

It has been shown\cite{ding} that, on the square lattice, the
antiferromagnetic Heisenberg model has a correlation length of order
$\xi(T)/a \sim 1$ for $T = J$ and of order $\xi(T)/a \sim
30$ for $T= 0.3 J$. 
On the other hand for the antiferromagnetic Heisenberg model on the
isotropic triangular lattice, the correlation length is only of order
a lattice constant at $T= 0.3J$.\cite{elstner} 
Thus the correlation length, $\xi(T)/a = 3.5\pm2.5$, obtained from the
analysis of the data for \kbr~is reasonable and places the materials
between the square and isotropic antiferromagnetic Heisenberg model as
has been argued on the basis of electronic structure
calculations.\cite{mckenzie:comments,kino,powell:review}

One of the best ways to measure antiferromagnetic correlation length is
by inelastic neutron scattering experiments. To perform this
experiment, one needs high quality single crystal. Unfortunately, it is
difficult to grow sufficiently large single crystals for \kpx; however,
recently some significant progress has been made.\cite{taniguchi-cry}
Another way to probe the correlation length is through the spin echo
experiment. The spin echo decay rate $1/T_2$ is proportional to the
temperature dependence correlation length [see Eq. (\ref{nmr_af})] so
measurements of $T_2$ would give us direct knowledge on the nature of
the correlation length. To the authors' knowledge there is no spin echo
decay rate measurement on the metallic phase of the layered organic
materials at the present time thus it is very desirable to have
experimental data on $T_2$ measurement to compare with the value of
$\xi(T)$ we have extracted above.

\subsection{The Knight Shift}\label{knight-shift}

As we pointed out in Section II the Knight shift $K_s$ will generally
have a weak temperature dependence throughout the whole temperature
range and so, thus far, we have neglected its temperature dependence.
However, it is apparent from Eq. (\ref{nmr_af2}) that for any choice of
parameter values $\{\beta, \xi(T_x)/a$, and $T_x\}$, $K_s$ will always
increase monotically as the temperature decreases. Therefore the
temperature dependence of the Knight shift potentially provides an
important check on the validity of the spin fluctuation model. However,
in the following discussion one should recall the caveats (discussed in
section \ref{sect:sf}) on the validity of the calculation of the Knight
shift stemming from the assumption that the dynamics of the long
wavelength part of dynamical susceptibility relax in the same manner as
a Fermi liquid.

\begin{figure*}
\epsfig{file=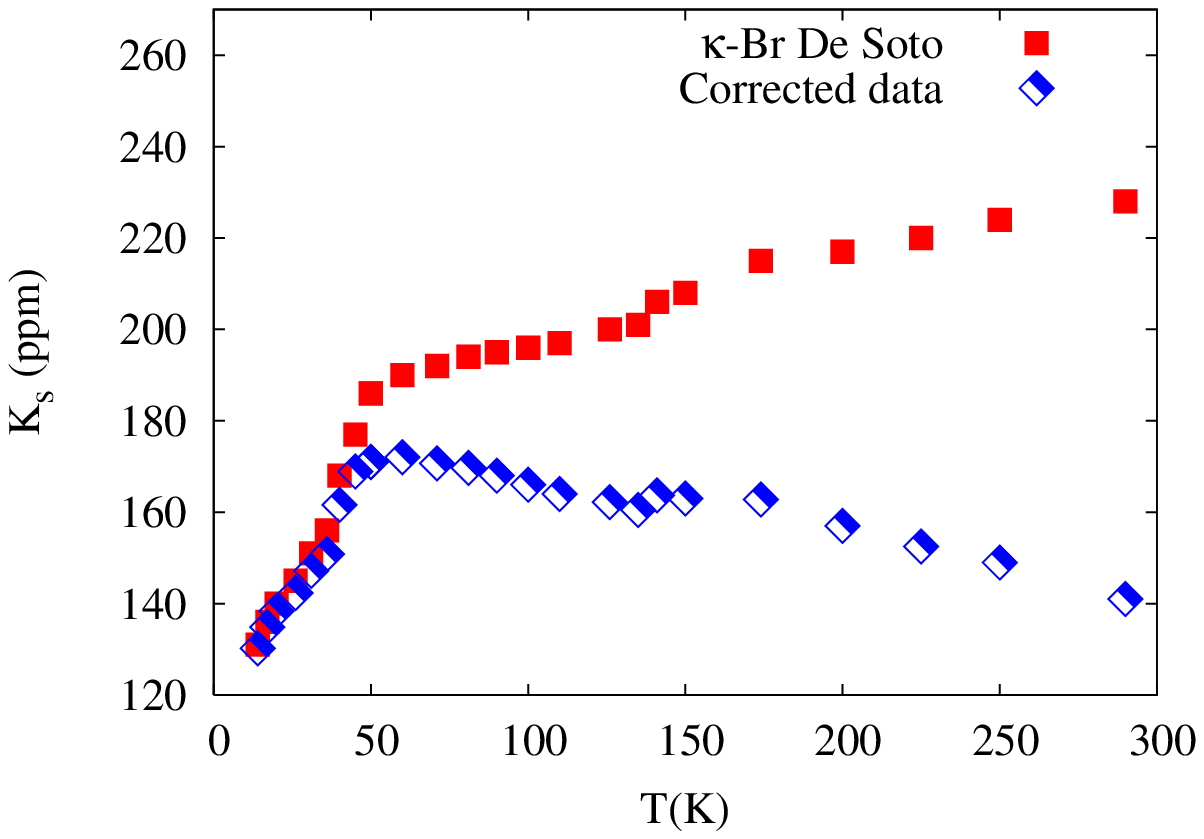, scale=0.7} \caption{The temperature
dependence of the (constant pressure) Knight shift as measured by
\kbr~by De Soto {\it et al}.\cite{desoto} (filled squares) and the
corrected Knight shift obtained by taking into account thermal
expansion of the lattice (half-filled diamonds), i.e., the constant
volume Knight shift. The temperature at which the Knight shift
decreases rapidly is about the same temperature at which $1/T_1T$ is
suppressed (see Fig. \ref{fig:t1t_fit2}), i.e. $T_\ks \sim T_\nmr$.
This suggests that $1/T_1T$ and $K_s$ are suppressed by the same
physics. In the limit of large correlation lengths, the spin
fluctuation model, Eq. (\ref{nmr_af2}), predicts a slowly varying
Knight shift which is almost temperature independent (solid line). The
discrepancy between theory and experiment arises because the model
calculates constant volume Knight shift while the experiment measures
constant pressure Knight shift.\cite{wzietek} Since \kpx~is soft there
will be a sizeable effect on $K_s$ due to the large thermal expansion.
The correction to the experimental Knight shift, by taking into account
these effects, was calculated by using Eq. (\ref{correction}) (half
filled squares). However, we stress that the lack of compressive
measurements of pressure and temperature dependence of the Knight
shift, isothermal compressibility, and thermal expansion means that
this correction is no better than an order of magnitude estimate.
However, our estimate indicates that there correction is large enough
that the data above $T_\ks$ cannot be shown to be in disagreement with
the spin fluctuation model. Below $T_\ks$ there is a clear disagreement
between the theory and data, in section \ref{phase} we argue that this
is because a pseudogap opens at $T_\nmr$. In the key to this figure
\kbr~is abbreviated as \kpbr.} \label{fig:correction}
\end{figure*}

In contrast to the prediction of the spin fluctuation model the
experimental data, e.g., those measured\cite{desoto} on
\kbr~(reproduced in Fig. \ref{fig:correction}), show that $K_s$
decreases slowly with decreasing temperature which then undergoes a
large suppression around $T_\ks \sim 50$ K. It should be emphasized
here that $T_\ks$ is approximately the same as $T_\nmr$, the
temperature at which $1/T_1T$ is maximum.

Since it is not possible to explain any of the NMR data below $T_\nmr$
in terms of the spin fluctuation model within the approximations
discussed thus far we focus on the temperature range between 50 K to
300 K just as we did for the analysis of $1/T_1T$. Even in this
temperature range, there is a puzzling discrepancy between theory and
experiment: the experimental data decreases slowly with decreasing
temperature while the theoretical calculation predicts the opposite. We
will argue below that this discrepancy arises because the data are
obtained at constant pressure while the theoretical prediction assumes
constant volume. Since the organic charge transfer salts are
particularly soft, thermal expansion of the unit cell may produce a
sizeable effect to the Knight shift and may not be neglected.

Following Wzietek {\it et al.},\cite{wzietek} we attempt to make an
estimate on the correction of the Knight shift for \kbr~due to
thermal expansion. Let us define $\chi_s^v(T,V)$ as the constant
volume spin susceptibility as a function of temperature. The
measured susceptibility is then given by a constant pressure
susceptibility $\chi_s^p = \chi_s^p[T,V(T,P)]$ while the theoretical
susceptibility is given by a constant volume susceptibility
$\chi_s^v = \chi_s^v[T,V(T=0,P)]$. The correction to the spin
susceptibility is then given by
\begin{eqnarray}
\Delta \chi &=& \chi_s^p-\chi_s^v
\\&=&\nonumber\int_0^T{dT'}\left(\frac{\partial \chi_s^p}{\partial
P}\right)_{T'} \left(\frac{V\partial P}{\partial V}\right)_{T'}
\left(\frac{\partial V}{V\partial T'}\right)_P.
\end{eqnarray}
The Knight shift is directly proportional to the spin susceptibility,
Eq. (\ref{ks}), which allows us to write the correction to the Knight
shift as
\begin{eqnarray}
\label{correction} \Delta K_s &=&
K_s^p-K_s^v\\&=&\nonumber\int_0^T{dT'}\left(\frac{\partial
K_s^p}{\partial P}\right)_{T'} \left(\frac{V\partial P}{\partial
V}\right)_{T'} \left(\frac{\partial V}{V\partial T'}\right)_P,
\end{eqnarray}
where $K_s^p$ is the (experimentally obtained) isobaric Knight shift,
$K_s^v$ is the (calculated) constant volume Knight shift, $(V
\partial P/\partial V)_T$ is the isothermal compressibility, and
$(\partial V/V \partial T)_P$ is the linear thermal expansion. It is
hard to obtain an accurate estimate for $\Delta K_s$ because there
are no complete sets of data for $K_s^p$, isothermal
compressibility, and thermal expansion as a function of temperature
and pressure for the \cation-X family. However a rough estimate for
$\Delta K_s$ may be made using the available experimental data. 

In Appendix \ref{sect:estimation} we estimate that 
\begin{eqnarray}
\left(\frac{\partial K_s^p}{\partial P}\right)_T &\sim& -3\times10^{-8}~\textrm{bar}^{-1}, \nonumber\\
\left(V \frac{\partial P}{\partial V}\right)_T &\sim& -10^5~\textrm{bar}, \nonumber\\
\nonumber \textrm{and}~ \left(\frac{1}{V}\frac{\partial V}{\partial
T}\right)_P &\sim& 10^{-4}~\textrm{K}^{-1}.
\end{eqnarray}
Combining these order of magnitude estimates we are able to obtain a
rough estimate on $\Delta K_s$ which can be written as $K_s^v \approx
K_s^p - 0.3 T$ for $T$ in Kelvin. The result is plotted in Fig.
\ref{fig:correction}. It is clear from the figure that our rough
estimate has already produced a non trivial correction to the Knight
shift. The Knight shift changes from having a positive slope in the raw
data to exhibit a rather small negative slope between $T_\ks \sim$ 50 K
and room temperature when the corrections to account for the thermal
expansion are included. The correction becomes small below about 50 K.
The lattice expansion clearly has a significant effect on the measured
Knight shift. To remove this effect one would need to either measure
the Knight shift at constant volume or pursue an experiment in which
the pressure dependence of $K_s$, isothermal compressibility, and
thermal expansion [c.f. Eq. (\ref{correction})] are measured
simultaneously to accurately determine $\Delta K_s$. Given the large
uncertainty in $\Delta K_s$ we take $K_s$ to be constant for
temperatures above 50~K in the rest of this paper. This is clearly the
simplest assumption, it is not (yet) contradicted by experimental data,
and, perhaps most important, any temperature dependence in the Knight
shift is significantly smaller than the temperature dependence of
$1/T_1T$.

Regardless of the valuse of $\Delta K_s$, the Knight shift calculated
from the spin fluctuation model is inconsistent with the experimental
data below $T_\ks \sim 50$ K (see Fig. \ref{fig:correction}). The
calculated $K_s$ shows a weakly increasing $K_s$ with decreasing
temperature, while the measured $K_s$ is heavily suppressed below 50 K.
One important point to emphasize here is that the temperature
dependence of $K_s$ will not change even if one uses the fully
q-dependent $A({\bf q})$ since $K_s$ [see Eq. (\ref{ks})] only probes
the ${\bf q}=\bf0$ component of the hyperfine coupling and
susceptibility. Thus, putting an appropriate q-dependent hyperfine
coupling will not change the result for $K_s$ (although it might give a
better description for $1/T_1T$). This provides a compelling clue that
some non-trivial mechanism is responsible to the suppression of
$1/T_1T$, $K_s$, and ${\cal K}$ below 50 K.

We have not addressed how the nuclear spin relaxation rate is modified
by the thermal expansion of the lattice. Since the organic compound is
soft, it is interesting to ask if there is a sizeable effect to
$1/T_1T$. Wzietek {\it et al.}\cite{wzietek} have performed this
analysis on quasi-1D organic compounds whose relaxation rate in found
to scale like $\chi_s^2$. One can straightforwardly derive the effect
of volume changes from the Hubbard model. If one uses the relation
$1/T_1T \sim \chi_s^2$ and assumes fixed $U$ and $t$, then $1/T_1T \sim
1/V^2$ will follow. However, it is clear from the phase diagram of the
organic charge transfer salts (Fig. \ref{fig:new_phase} and Ref.
\onlinecite{powell:review}) that there is a rather large change in $U$
and $t$ for even small pressure variations. Therefore, there is no
obvious relationship between $1/T_1T$ and $\chi_s$ for the quasi-2D
organics and it is not clear how the imaginary part of the
susceptibility $\chi''({\bf q},\omega)$, which enters $1/T_1T$, is
effected by thermal expansion and lattice isothermal compressibility.
More detailed experiments are clearly needed to determine the effect of
thermal expansion of the lattice on the measured relaxation rate.

\section{Spin Fluctuations in The Mott Insulating Phase of \kcn3}

Recent experiments on \kcn3~by Shimizu and
collaborators\cite{kawamoto:kcn3,shimizu,kurosaki,shimizu:prb2006} have
generated a lot of
interest.\cite{powell:review,ben:d+id,powell:group,zheng,motrunich,leelee,leelee2}
This is because the Mott insulating phase of this material appears to
have a spin liquid ground state, that is a state which does not have
magnetic ordering (or break any other symmetry of the normal state)
even though well-formed local moments exist. This is very different
from the Mott insulating phases of the other $\kappa$ salts, such as
\kcl, which clearly shows antiferromagnetic ordering\cite{miyagawa:kcl}
at low temperature and ambient pressure. An elegant demonstration of
these two different ground states is provided by susceptibility
measurements:\cite{shimizu} the susceptibility of \kcl~exhibits an
abrupt increase around 25 K which marks the onset of N\'eel ordering
while the susceptibility of \kcn3~shows no sign of a magnetic
transition. The transition to a magnetically ordered ground state
realized in \kcl~is also demonstrated by the splitting of NMR spectra
below the transition temperature.\cite{miyagawa:chemrev} The difference
in the ground states of \kcl~and \kcn3~appears to be connected with the
fact that there is significantly greater frustration in \kcn3~(for
which $t'/t\sim1$) than there is in \kcl~ (for which $t'/t\sim0.7$).
Geometrical frustration alone is not sufficient to explain the absence
of magnetic order in \kcn3~because a Heisenberg model on an isotropic
triangular lattice is known to exhibit a magnetically ordered ground
state, i.e. 120$^\circ$ state. It may be that the proximity to the Mott
transition plays an important role in allowing the absence of magnetic
ordering at low temperatures in \kcn3. One possible explanation for the
existence of a spin liquid ground state is there are ring exchange
terms in the Hamiltonian arising from charge fluctuations which has
been studied by several groups.\cite{motrunich,leelee,leelee2}

\begin{figure*}
\begin{centering}
\hspace*{12.3cm}\epsfig{file=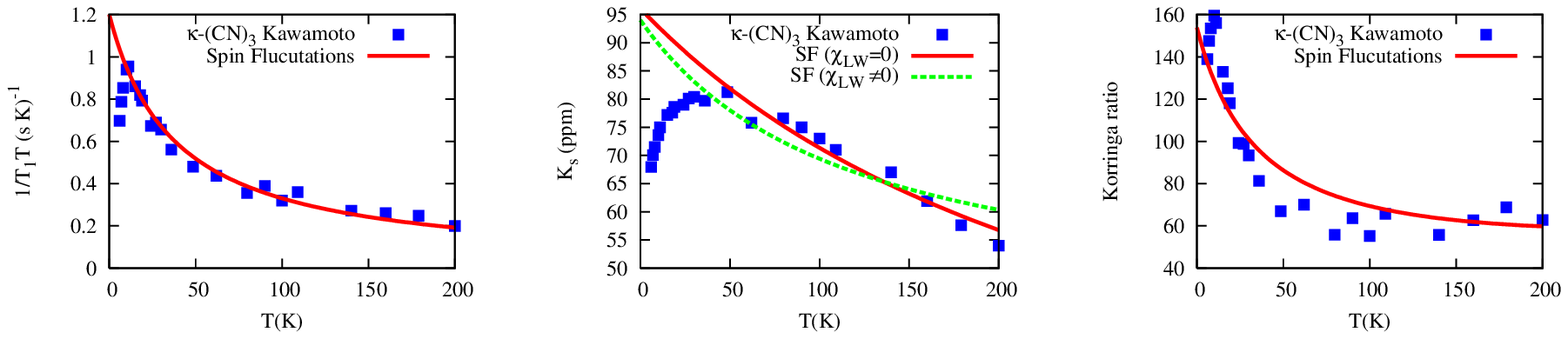, scale=0.98}
\end{centering}
\caption{[Color online] Comparison of the spin fluctuation theory with
the measured\cite{kawamoto:kcn3} temperature dependence of the nuclear
spin relaxation rate per unit temperature, $1/T_1T$ (left panel),
Knight shift, $K_s$ (center panel), and Korringa ratio (right panel) of
the Mott insulating phase of \kcn3~[abbreviated as \kpcn3~in the keys
to the figures]. The spin fluctuation model is in good agrement with
the measured $1/T_1T$, but does not describe the Knight shift (and
hence $\cal K$) well. We have also checked that using the Fermi liquid
for $\chi_\lw$ does not improve the fit to the $K_s$ data, and this fit
is shown as a dashed line. This fit to the data is clearly worse than
simply setting $\chi_\lw=0$. Also note that the peak in $1/T_1T$,
$T_\nmr$, is at a lower temperature than the maximum in the Knight
shift, $T_\ks$. This behavior is qualitatively different from that of
the other $\kappa$ salts where $T_\nmr\sim T_\ks$ (see Figs.
\ref{fig:t1t_fit2} and \ref{fig:correction}). This suggests that in
\kcn3~the origin of the $1/T_1T$ suppression is different from physics
that gives rise to the maximum in $K_s$. The parameter values from the
lines of best fit shown in this figure are reported in Table
\ref{tab:parameter2}. The fact that the model gives a reasonably good
fit to $1/T_1T$ but not to the $K_s$ data suggests that the spin
fluctuation model fails to correctly account for the long wavelength
physics, but suggests that the model correctly describes the physics
around a peak in $\chi({\bf q},\omega)$ which dominates the integral
over the first Brillouin zone and thus $1/T_1T$ [c.f., Eq.
(\ref{t1t})]. This is rather surprising as the correlation length is
less than one lattice constant at $T=50$~K. This result is clearly
inconsistent with the initial assumption that the long wavelength
susceptibility is dominated by a peak in the dynamic susceptibility at
a finite wave vector. Finally we note that the fact that ${\cal K}>1$
shows that the spin fluctuations are antiferromagnetic, this is rather
interesting given the importance of Nagaoka ferromagnetism on the
triangular lattice.\cite{nacoo}} \label{fig:kcn3}
\end{figure*}

The NMR relaxation rate in \kcn3~(Ref. \onlinecite{kawamoto:kcn3} and
Fig. \ref{fig:kcn3}) shows a similar temperature dependence to that in,
for example, \kbr. $1/T_1T$ is enhanced over the Korringa-like behavior
with a peak at $T_\nmr \sim 10$ K below which it exhibits a large
decrease. However the Knight shift $K_s$ in \kcn3~is quite different to
that in \kbr~(compare Figs. \ref{fig:correction} and \ref{fig:kcn3}).
In \kcn3, $K_s$ increases as the temperature is lowered from room
temperature until it reaches a broad maximum around $T_\ks \sim 30-50$
K below which it drops rapidly. In contrast, $K_s$ in \kbr~shows a weak
temperature dependence down to $T_\ks$ below which it undergoes a sharp
decrease (see Fig. \ref{fig:correction}). Another difference is
$T_\nmr$ is considerably lower than $T_\ks$ in \kcn3~whereas they are
roughly the same in \kbr. This suggests that whereas the suppression of
$1/T_1T$ below $T_\nmr$ and $K_s$ below $T_\ks$ in \kbr~probably has a
common origin; in \kcn3~the origin of the suppression of $1/T_1T$ below
$T_\nmr$ is different from the origin of the broad maximum in $K_s$ at
$T_\ks$. Note that the fact that ${\cal K}>1$ shows that the spin
fluctuations are antiferromagnetic, this is rather interesting given
the importance of Nagaoka ferromagnetism on the triangular
lattice.\cite{nacoo}

Given the reasonable agreement between the antiferromagnetic spin
fluctuation model with the NMR data on \kbr~(down to $T_\nmr \sim 50$
K), we apply the same formalism to \kcn3. A slight modification to the
spin fluctuation model is necessary since, unlike the other $\kappa$
salts studied in this paper, \kcn3~is an insulator. Therefore we
clearly cannot use the Fermi liquid form of $\chi_\lw$. The simplest
approximation is that the dynamic susceptibility given in Eq.
(\ref{dynamic}) will only consist of $\chi_\af({\bf q},\omega)$. In the
region where $\xi(T)/a$ is large, $\chi_Q =\alpha (\xi/a)^{2-\eta}$ and
$\omega_\sf = \alpha' (\xi/a)^{-z}$ where $\alpha$ and $\alpha'$ are
temperature independent constants and $a$ is the lattice spacing.
Within these approximations the nuclear spin relaxation rate and Knight
shift are given by
\begin{eqnarray}
\frac{1}{T_1T} &=& \left(\frac{1}{T_1T}\right)_0
\frac{(\xi/a)^{2+z-\eta}}{1+(Qa)^2(\xi/a)^2}\nonumber\\
K_s &=& (K_s)_0 \frac{(\xi/a)^{2-\eta}}{1+(Qa)^2(\xi/a)^2}
\label{nmr_kcn3}
\end{eqnarray}
with
\begin{eqnarray}
\left(\frac1{T_1T}\right)_0&=&\frac{2\alpha k_B |A|^2 }{\alpha'\gamma_e^2\hbar^4}\nonumber\\
(K_s)_0&=&\frac{\alpha'|A|}{\gamma_e \gamma_N \hbar^2},
\label{coefficient_kcn3}
\end{eqnarray}
where ${\bf Q}=(Q,Q)$ is the finite wave vector on which we assume
the susceptibility to peak. Again, we take the temperature
dependence of the correlation length to be $\xi(T)/\xi(T_x) =
\sqrt{2T_x/(T+T_x)}$.

Following the same approximation scheme as before (outlined in Section
II B), we assume a relaxational dynamics of the spin fluctuations,
which are described by a dynamic critical exponent $z=2$, and a mean
field critical exponent $\eta=0$. Within these approximations
$\omega_\sf = \alpha (a/\xi)^{2}$ and $\chi_Q =\alpha' (\xi/a)^2$. The
nuclear spin relaxation rate and Knight shift are then given by
\begin{eqnarray}
\frac{1}{T_1T} &=& \frac{(1/T_1T)_0 C^2}
{(T/T_x+1)^2+(Qa)^2C(T/T_x+1)}\nonumber\\
K_s &=& \frac{(K_s)_0 C }{1+(Qa)^2C+T/T_x}. \label{nmr_af_kcn3}
\end{eqnarray}
We work in the high temperature approximation for $1/T_1T$ - following
a procedure similar to that employed to obtain Eq.
(\ref{limiting_t1t}). If the dynamic susceptibility is strongly peaked
at $q=Q$ then the parameter $2(Qa)^2(\xi(T_x)/a)^2$ is much larger than
1 so $(Qa)^2C(T/T_x+1)$ is larger than $(T/T_x+1)^2$ and we can just
keep the term proportional to $(Qa)^2$ in the denominator which allows
us to write $1/T_1T$ as
\begin{equation}
\frac{1}{T_1T} \simeq
\frac{(1/T_1T)_0[\sqrt{2}\xi(T_x)/a]^2}{(Qa)^2(T/T_x+1)}
\label{limiting_t1t_kcn3}
\end{equation}
We use the expressions for $1/T_1T$ given in Eq.
(\ref{limiting_t1t_kcn3}) and for $K_s$ in Eq. (\ref{nmr_af_kcn3}) to
fit the \kcn3~data.\cite{kawamoto:kcn3} We assume the susceptibility
has a strong peak at $Q=2\pi/3$.\footnote{Note that the location of the
peak does not qualitatively effect the theory, unless it is at $Q=0$.
If the peak is elsewhere then it will simply effect the magnitude of
$\xi(T_x)$.} The parameters of the best fit are reported in Table
\ref{tab:parameter2} and the results are plotted in Fig \ref{fig:kcn3}.
While the spin fluctuation model produces a reasonably good fit to the
$1/T_1T$ above $T_\nmr \sim 10$ K, it does not reproduce $K_s$ data
well as can be seen from the upward curvature in the fit in contrast to
the data which shows a slight downward curvature. We also performed the
fit to \kcn3~data using the most general forms Eq. (\ref{nmr_kcn3}) and
taking $z$ and $\eta$ as free parameters. Good fits to $1/T_1T$ and
$K_s$ can be obtained with $\eta \sim 1$ and $z \sim 2$ but this gives
us so many free parameters that the value of such fits must be
questioned.

An important fact is that $K_s$ probes the long wavelength dynamics. We
have set $\chi_\lw (\omega)$ to zero in order to make the simplest
possible assumption about the insulating state. The data indicate that
this assumption is probably incorrect. Recently Zheng {\it et
al.}\cite{zheng} used a high temperature series expansion to calculate
the uniform spin susceptibility (which is the same as the Knight shift
apart from a constant of proportionality) for the Heisenberg model on a
triangular lattice, applied it to \kcn3. They obtained a good agreement
with the experimental data. The spin fluctuation model described here
can be
viewed as a different route to understand the same experiment. 
The discrepancy between the spin fluctuation model and the data
suggests a failure of our implicit assumption that the long wavelength
physics, which determines $K_s$, is dominated by a peak in the dynamic
susceptibility due to spin fluctuations. This is consistent with the
fact that we find that  $\xi(T) \sim 0.2-0.4$ lattice spacings at
$T=50$ K which clearly disagrees with our initial assumption that
$\xi(T) \gg a$. This begs the question what physics dominates the long
wavelength physics both in the series expansions and in the real
material?

\begin{table}
\begin{tabular}{c | ccc c}
\hline\hline
Parameter & & & Fit results\\
\hline
$(1/T_1T)_0$ $(s^{-1}$K$^{-1}$) & & & 220 $\pm 11$ \\
$(K_s)_0$& & & 8100 $\pm$ 720 \\
$T_x$ (K)& & & 40 $\pm$ 4 \\
$\xi(T_x)/a$ & & & 0.3 $\pm$ 0.1\\
\hline\hline
\end{tabular}
\caption{The parameters obtained from the best fits to $1/T_1T$ and
$K_s$ data in \kcn3. These parameters are used to produce Fig.
\ref{fig:kcn3}. The antiferromagnetic correlation length $\xi(T_x)/a$
is short ranged consistent with the significant
frustration\cite{shimizu,zheng,zheng2} present in this material.}
\label{tab:parameter2}
\end{table}

The low temperature properties of \kcn3~are clearly inconsistent with a
magnetic ordered ground state. For a two-dimensional quantum spin
system with an ordered ground state, the low temperature properties are
captured by the non-linear sigma model. The observed temperature
dependence of $1/T_1$ and the spin echo rate $1/T_2$ follow $1/T_1
\propto T^{7/2} \xi(T)$, and $1/T_2 \propto T^{3} \xi(T)$, where the
correlation length $\xi(T)$ is given by\cite{chubukov}
\begin{equation}
\frac{\xi(T)}{a} = 0.021\left(\frac{c}{\rho_s}\right) \left(\frac{4\pi
\rho_s}{T}\right)^{1/2} \exp\left(\frac{4\pi \rho_s}{T}\right)
\end{equation}
where $c$ is the spin wave velocity and $\rho_s$ is the spin stiffness.
In the quantum critical regime,\cite{chubukov} $1/T_1T \sim T^{\eta-1}$
and $1/T_2 \sim T^{\eta-1}$ [c.f., Eq. (\ref{nmr_qc})] where $\eta$ is
the anomalous critical exponent associated with the spin-spin
correlation function whose value is generally less than 1. Thus for a
magnetically ordered state, which can be well described by O(N) non
linear sigma model, both $1/T_1T$ and $1/T_2$ should \emph{increase}
with decreasing temperature. For \kcn3~Shimizu {\it et
al}.\cite{shimizu2} found that $1/T_1T \sim T^{1/2}$ and $1/T_2 \sim $
constant from 1 K down to 20 mK which suggests the critical exponent
$\eta > 1$. The nuclear spin relaxation rate {\it decreases} with
decreasing temperature. Such a large value of $z$ is what occurs for
deconfined spinons.\cite{chubukov}

\section{Unconventional Coherent Transport Regime and the \kpx~Phase
Diagram\label{phase}}

In Section \ref{sect:dmft}, we discussed the DMFT description of the
crossover from a bad metal to a  Fermi liquid. DMFT successfully
predicts the unconventional behaviors observed  in a number of
experiments on the \kpx~salts. These include the resistivity,
thermopower, and ultrasound velocity. The unconventional behaviors seen
in these measurements are associated with the crossover from bad
metallic regime to a renormalized Fermi liquid in the DMFT picture.
While DMFT gives reliable predictions for the transport
properties,\cite{Jaime-HH-DMFT,merino,limelette,hassan} it is not able
to explain the loss of DOS observed in the nuclear spin relaxation rate
and Knight shift. Thus the NMR data suggest that the coherent transport
regime is not simply a Fermi liquid, contrary to what has previously
been thought.

\begin{figure*}
\epsfig{file=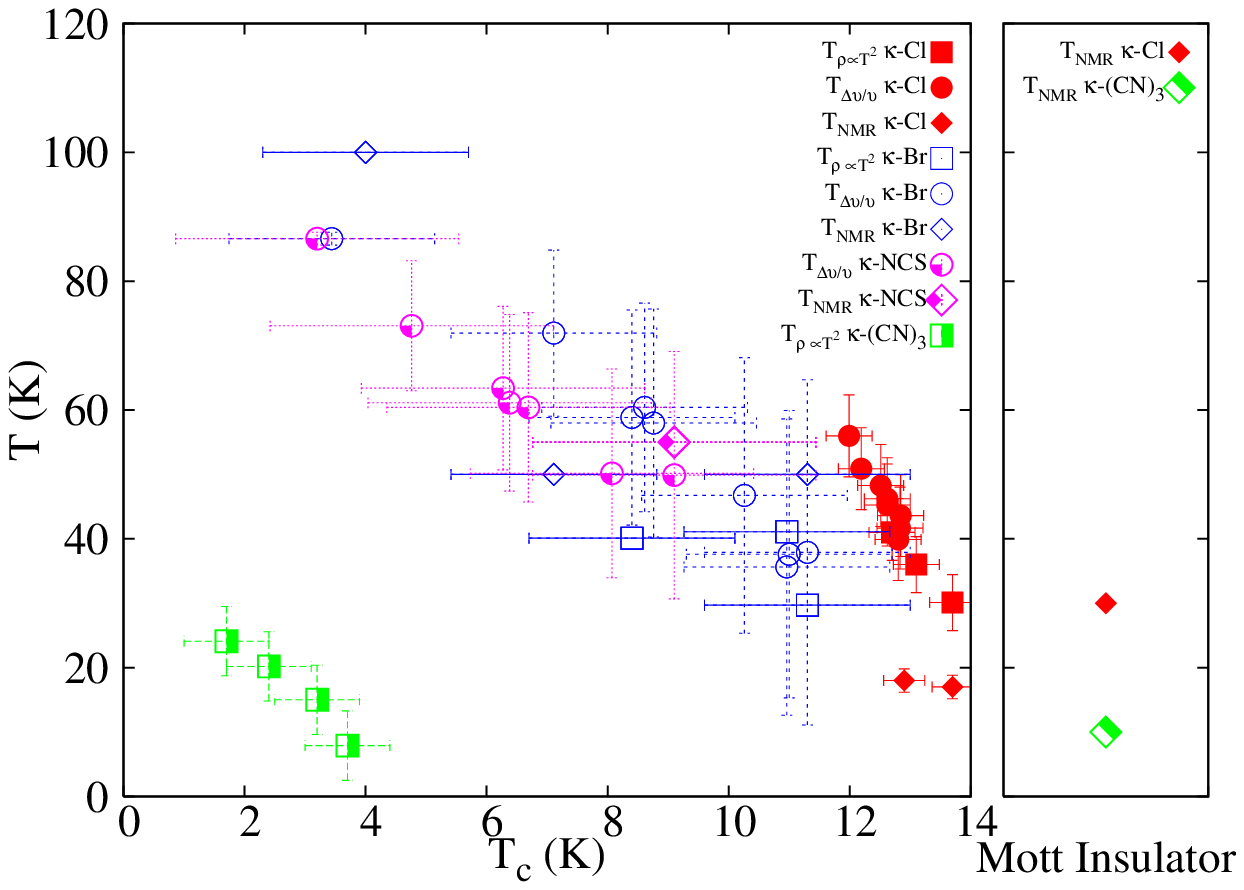,scale=1.0} \caption{[Color online] The
relationship between different temperature scales for a range of
organic charge transfer salts. The superconducting transition
temperature $T_c$ is used to parameterize the proximity of the material
to the Mott transition ($T_\c$ decreases as one moves further away from
the Mott transition). A plot of $T_\res, T_\us,$ and $T_\nmr$ against
$T_\c$ for several \kpx~ salts shows that the peak in $1/T_1T$,
$T_\nmr$, occurs at the same temperature as the crossover form a bad
metal to the coherent transport regime measured in transport ($T_\res$)
and ultrasonic attenuation ($T_\us$) experiments. $T_\res$ is the
temperature at which the resistivity deviates from a $T^2$ behavior,
$T_\us$ is the temperature at which a dip in the ultrasound velocity is
observed. The left panel shows the data for the \kpx~family in the
metallic phase \{\kcl~and \kcn3~under pressure, \kbr~and \kncs\} while
the right panel shows the data for the \kpx~family in the insulating
phase [\kcl~and \kcn3~at ambient pressure]. In the metallic phase we
use $T_c$ as a single parameter to characterize the effect of chemical
substitution and hydrostatic pressure. This works surprisingly well and
the data for \kcl, \kbr, and \kncs~ is seen to collapse roughly onto a
single trend, which suggests that the spin fluctuations in the metallic
phases are rather similar. In contrast, the data for \kcn3~fall onto a
separate curve which suggests that there are important differences
between the spin fluctuations in this material and those in other
$\kappa$ phase salts. This is perhaps not so surprising in light of the
fact that \kcn3~has a spin liquid round state in its Mott insulating
phase while the materials are close to at N\'eel ordered Mott
insulating phase. This plot suggests that the pseudogap opens at the
same temperature as the crossover from bad metal to coherent transport
regime. Whether this is because of a deep link between the crossover
and the pseudogap or because the lack of coherence in the bad metal
destroys the pseudogap remains to be seen. Collectively these data show
that the coherent transport regime is not simply a renormalised Fermi
liquid as has previously been thought. It should be emphasized that the
large error bars are the result of our estimates of the systematic
errors produced by equating pressures in different experiments. The
procedure to obtain the error bars presented in this plot is discussed
in Appendix \ref{sect:errors}. The symbols represent both the material
and the experiment as follows: filled symbols correspond to the data
for \kcl, open symbols denote \kbr, open symbols with black dots denote
\kncs, and half filled symbols denote \kcn3. The square symbols
represent $T_\res$ vs $T_\c$, circles represent $T_\us$ vs $T_\c$, and
triangles represent $T_\nmr$ vs $T_\c$. The references from which the
data were collected are given in Table \ref{tab:ref}.}\label{fig:phase}
\end{figure*}

\begin{table}
\begin{tabular}{c | c | c | c | c}
\hline\hline
Material & ~$T_\c(P)$~ & ~$T_\res(P)$~ & ~$T_\us(P)$~ & ~$T_\nmr(P)$\\
\hline
\kpcl & [\onlinecite{lefebvre,fournier}] & [\onlinecite{limelette}] & [\onlinecite{fournier}] & [\onlinecite{lefebvre}]\\
\kpbr & [\onlinecite{frikach,schirber}] & [\onlinecite{strack}] & [\onlinecite{frikach}] & [\onlinecite{mayaffre}]\\
\kpncs & [\onlinecite{caulfield}] & - & [\onlinecite{frikach}] & [\onlinecite{kawamoto}]\\
\kpcn3 & [\onlinecite{kurosaki}] & [\onlinecite{kurosaki}] & - & [\onlinecite{kawamoto:kcn3}]\\
\hline\hline
\end{tabular}
\caption{The references from which the pressure dependence of different
temperature scales for different materials used to produce Fig.
\ref{fig:phase} were taken. The notation is the same as that given in
Fig. \ref{fig:phase}. $T_\res$ is the temperature at which the
resistivity deviates from a $T^2$ behavior, $T_\us$ is the temperature
at which a dip in the ultrasound velocity is observed, and $T_\nmr$ is
the temperature on which $1/T_1T$ peaks. In the table \kpcl, \kpbr, and
\kpncs~and \kpcn3~are used as shorthand for \kcl, \kbr, \kncs, and
\kcn3~respectively.}\label{tab:ref}
\end{table}

To illustrate the nature of the low temperature paramagnetic metallic
state more qualitatively, it is instructive to study how the bad
metal-coherent transport crossover is related to the loss of DOS.
Therefore we have investigated the relationship between $T_\res$, the
temperature at which the resistivity deviates from $T^2$ behavior;
$T_\us$, the temperature at which a dip in the ultrasonic velocity is
observed; and $T_\nmr$, the temperature at which $1/T_1T$ (and $K_s$)
is maximum which appears to mark the onset of a loss of DOS. In Figure
\ref{fig:phase} we plot $T_\res$, $T_\us$, and $T_\nmr$ as measured by
several different groups for various salts against $T_\c$ which serves
well as a single parameter to characterize both the hydrostatic
pressure and the variation in chemistry (or `chemical pressure'). This
analysis is complicated by the necessity of comparing pressures from
different experiments. Our procedure for dealing with this issue is
outlined in the Appendix \ref{sect:errors}, and we stress that the
large error bars in Fig. \ref{fig:phase} are due to the difficulties in
accurately measuring pressure rather than problems in determining
$T_c$, $T_\res$, $T_\us$, or $T_\nmr$.

It is clear from Fig. \ref{fig:phase} that the data for \kcl, \kbr, and
\kncs~ fall roughly onto a single curve. This suggests that $T_\nmr$
coincides with $T_\res$ and $T_\us$. Thus the loss of DOS, associated
with $T_\nmr$, occurs around the temperature at which the crossover
from bad metal to coherent transport regime takes place. The loss of
DOS observed in $1/T_1T$ and $K_s$ is not what one would expect for a
Fermi liquid; therefore the coherent intralayer transport regime is
more complicated than a renormalized Fermi liquid. This must result
from non-local correlations which are not captured by DMFT since DMFT
captures local correlations exactly. One possible explanation for the
loss of DOS is  the opening of a pseudogap.

Another important point to emphasize from Fig. \ref{fig:phase} is the
appearance of a second trend formed by \kcn3~which is clearly distinct
from the trend of the data points from \kcl, \kbr, and \kncs. This
shows that the spin fluctuations in \kcn3~are qualitatively different
from those in the other \kpx~salts. Of course, qualitative differences
are not entirely unexpected due to the spin liquid rather than
antiferromagnetic ground state in \kcn3.  It has recently been argued
that the differences in the spin fluctuations in \kcn3~will lead this
material to display a superconductivity with a different symmetry of
the order parameter than the other
\kpx~salts.\cite{ben:d+id,powell:group} This result shows that the spin
fluctuations are indeed qualitatively different in \kcn3.

On the basis of the above analysis we sketch the phase diagram of \kpx,
shown in Fig. \ref{fig:new_phase}. The pseudogap phase shows an
interesting set of behaviors. On one hand it exhibits a loss of DOS as
is evident from $1/T_1T$ and $K_s$. On the other hand, it exhibits
coherent intralayer transport as is shown by the $T^2$ resistivity
behavior\cite{limelette}; it also has long lived quasiparticles and a
well defined Fermi surface clearly seen from de Haas-van Alphen and
Shubnikov-de Haas oscillation
experiments.\cite{wosnitza,singleton,kartsovnik} One framework in which
it may be possible to understand both of these sets of behaviors is if
there is a fluctuating superconducting gap.\cite{scarratt} This idea
has been applied to the cuprates;\cite{schmalian1,schmalian2} it would
be interesting to see whether such an approach gives a good description
of \kpx. Another interesting observation is that the measurements which
see the loss in the DOS probe the spin degrees of freedom whereas the
evidence for well defined quasiparticles comes from probes of the
charge degrees of freedom. This may be suggestive of a `spin gap' which
could result from singlet formation as in the RVB picture.

To date there have been few experiments studying the pressure
dependence of $1/T_1T$ or $K_s$. Therefore it is not possible, at
present, to determine with great accuracy where the pseudogap vanishes.
The available NMR experiment under
pressure\cite{mayaffre,wzietek-review} suggest that at sufficiently
high pressures the pseudogap onset temperature is lower than the
incoherent-coherent crossover temperature and pressure eventually
suppresses the pseudogap altogether. We represent the current
uncertainty over where the pseudogap is completely suppressed by
pressure by drawing a shaded area with a question mark in the phase
diagram. It is plausible that the pseudogap vanishes very close, if not
at the same pressure, to the point where the superconducting gap
vanishes. This would be consistent with RVB
calculations\cite{ben:prl,zhang} which suggest that the pseudogap and
superconducting gap are proportional to each other and so should vanish
at about the same pressure. However, we should stress that there really
is not yet sufficient data to determine exactly where the pseudogap
vanishes and admit that our choice is, perhaps, a little provocative.

Clearly a series of careful experiments are required to elucidate when
$T_\nmr$ tends to zero. Understanding where the pseudogap vanishes is
an important consideration in light of the number of theories based on
a hidden pseudogap quantum critical point in the
cuprates.\cite{Sachdev} Furthermore, the superconducting state in
organic charge transfer salts  far from the Mott transition are highly
unconventional. These low $T_c$ organic charge transfer salts have
unexpectedly large penetration depths\cite{ujjual,pratt} and are not
described by BCS theory.\cite{constraints} The possibility that a
quantum critical point is associated with the pressure where $T_c$ goes
to zero invites comparison with the heavy fermion material
CeCoIn$_{5-x}$Sn$_{x}$ \cite{CeCoIn} in which a  quantum critical point
seems to be associated with the critical doping to suppress
superconductivity. Thus low $T_c$ organic charge transfer salts appear
increasingly crucial for our understanding of the organic charge
transfer salts.\cite{powell:review}

An important question to address theoretically is why $T_\nmr$ might
coincide with $T_\res$ and $T_\us$. Of  course, it may be that the two
phenomena are intimately connected. However, another possibility
suggests itself on the basis of DMFT and RVB calculations. DMFT
correctly captures the local physics and it is this local physics that
dominates the cross-over from a `bad-metal' to coherent in-plane
transport. On the other hand RVB does not capture this cross-over
(because the Mott transition is only dealt with at the Brinkmann-Rice
level\cite{brinkman}) but does capture some of the non-local physics
which DMFT neglects. The pseudogap is predicted by RVB theory to
increase in temperature when pressure is lowered.\cite{ben:prl,zhang}
This rise in the pseudogap temperature is predicted to continue until
the pressure is lowered all the way to the Mott transition, in contrast
with the observed behavior (c.f., Figs \ref{fig:new_phase} and
\ref{fig:phase}). However, we conjecture that the RVB physics is `cut
off' by the loss of coherence at $T_\res$ and $T_\us$ thus preventing
$T_\nmr$ from exceeding $T_\res\approx T_\us$.

\section{Conclusions}

We have applied a spin fluctuation model to study the temperature
dependences of the nuclear spin relaxation rate and Knight shift in the
paramagnetic metallic phases of several quasi two-dimensional organic
charge transfer salts. The large enhancement of $1/T_1T$ between
$T_\nmr$ \{$\sim50$~K in \kbr\} and room temperature has been shown to
be the result of strong antiferromagnetic spin fluctuations. The
antiferromagnetic correlation length is estimated to be $3.5\pm2.5$
lattice spacings in \kbr~at $T=50$ K. The temperature dependence of
$1/T_1T$ for $T \> T_\nmr$ from the spin fluctuation model is
qualitatively similar with the predictions of DMFT. The spin
fluctuations in \kcl, \kbr, \deut8br, and \kncs~are found to be
remarkably similar both qualitatively and quantitatively. Strong spin
fluctuations seem to be manifested in materials close to Mott
transition. Recent NMR experiments\cite{kawamoto:ag} on \cation
Ag(CN)$_2\cdot$H$_2$O, which is situated further away from the Mott
transition, suggests that the spin fluctuations in this materials are
not as strong as those in the other $\kappa$ salts studied here.

We have also applied the spin fluctuation formalism to the strongly
frustrated system \kcn3. In this compound the measured $1/T_1T$, which
probes the entire Brillouin zone, agrees well with the predictions of
the spin fluctuation model for $T>T_\nmr\sim10$~K. In contrast the
measured Knight shift, which only depends on the long wavelength
physics, is not well described by the spin fluctuation model. This
suggests that at least one of the assumptions made in the spin
fluctuation model: (i) $z=2$, $\eta=0$, or (ii) $\chi({\bf q},\omega)$
is strongly peaked at wave vector ${\bf q}={\bf Q}$; is violated in
\kcn3~or that the model neglects some important long wavelength
physics. In light of the recent evidence for a spin liquid ground state
in \kcn3, in contrast to the antiferromagnetic or `$d$-wave'
superconducting grounds states in  \kcl, \kbr, \deut8br, and \kncs, and
the greater degree of frustration in \kcn3~it is interesting that there
are such important qualitative and quantitative differences between the
spin fluctuations in \kcn3~and those in the other $\kappa$ phase salts.

The peak of $1/T_1T$ and the suppression of $K_s$ are strongly
dependent on pressure: they are systematically reduced and completely
vanish at high pressure (\> 4 kbar);\cite{mayaffre} at high pressure a
Korringa-like temperature dependences of $1/T_1T$ and $K_s$ are
recovered for all temperatures. It is clear that high pressures will
suppress both the antiferromagnetic spin fluctuations which are
dominant above 50 K and the mechanism (presumably the pseudogap) which
causes drops in $1/T_1T$ and $K_s$ below 50 K at ambient pressure.


The large suppression of $1/T_1T$ and $K_s$ below $T_\nmr$ observed in
all the $\kappa$ salts studied here cannot be explained by the M-MMP
spin fluctuation model. The most plausible mechanism to account for
this feature is the appearance of a pseudogap which causes the
suppression of the density of states at the Fermi energy. This is
because at low temperature $1/T_1T$ and $K_s$ are proportional
$\tilde{\rho}^2(E_F)$ and $\tilde{\rho}(E_F)$, respectively [c.f., Eq.
(\ref{nmr_correlated})]. Independent evidence for the suppression of
density of states at the Fermi level comes from the linear coefficient
of specific heat $\gamma$.\cite{timusk} The electronic specific heat
probes the density of excitations within $k_B T$ of the Fermi energy.
Any gap will suppress the density of states near the Fermi surface
which results in the depression of the specific heat coefficient
$\gamma$. Kanoda\cite{kanoda:jspj2006} compared $\gamma$ for several of
the \kpx~salts and found that in the region close to the Mott
transition, $\gamma$ is indeed reduced. One possible interpretation of
this behavior is a pseudogap which becomes bigger as one approaches the
Mott transition. However, other interpretations are also possible, in
particular one needs to take care to account for the coexistence of
metallic and insulating phases; this is expected as the Mott transition
is first order in the organic charge transfer
salts.\cite{kagawa,sasaki} The existence of a pseudogap has also been
suggested $\lambda$-(BEDT-TSF)$_2$GaCl$_4$\cite{suzuki} from microwave
conductivity. The reduction of the real part of the conductivity
$\sigma_1$ from the Drude conductivity $\sigma_\mathrm{dc}$ and the
steep upturn in the imaginary part of the conductivity $\sigma_2$ may
be interpreted in term of preformed pairs leading to a pseudogap in
this material.

The experimental evidence from measurements of $1/T_1T$, $K_s$, and
heat capacity all seem to point to the existence of a pseudogap below
$T_\nmr$ in \kbr~and \kncs. Thus a phenomenological description which
takes into account both the spin fluctuations which are important above
$T_\nmr$ and a pseudogap which dominates the physics below $T_\nmr$
would seem to be a reasonable starting point to explain the NMR data
for the entire temperature range (clearly superconductivity must also
be included for $T<T_\c$). We will pursue this approach in our future
work. In particular one would like to answer the following questions:
how big is the pseudogap and what symmetry does it have? Is there any
relation between the pseudogap and the superconducting gap? The answer
to these questions may help put constraints on the microscopic
theories.

{\it Future experiments.} There are a number of key experiments to
study the pseudogap. The pressure and magnetic field dependences of the
nuclear spin relaxation rate and Knight shift will be valuable in
determining the pseudogap phase boundary, estimating the order of
magnitude of the pseudogap, and addressing the issue how the pseudogap
is related to superconductivity. In the cuprates, there have been
several investigations of the magnetic field dependence of the
pseudogap seen in NMR experiments. For
Bi$_2$Sr$_{1.6}$La$_{0.4}$CuO$_6$ the nuclear spin relaxation rate does
not change will field up to 43 T.\cite{gqzheng2} However, since $T^*
\sim 200$ K, one may require a larger field to reduce the pseudogap.
Similar results were found in YBa$_2$Cu$_4$O$_8$.\cite{gqzheng}
However, in YBa$_2$Cu$_3$O$_{7-\delta}$ [see especially Fig. 6 of Ref.
\onlinecite{mitrovic}] a field of order 10 T is enough to start to
close the pseudogap. Mitrovic {\it et al}.\cite{mitrovic} interpret
this observation in terms of the suppression of `$d$-wave'
superconducting fluctuations.

The interlayer magnetoresistance of the cuprates has proven to a
sensitive probe of the pseudogap.
\cite{mozorov,shibauchi,kawakami,elbaum} Moreover, it is found that for
the field perpendicular to the layers (which means that Zeeman effects
will dominate orbital magnetoresistance effects) the pseudogap is
closed at a field given by
\begin{equation}
H_{PG} \simeq \frac{\hbar k_B T^* }{ \gamma_e}
\end{equation}
where $T^*$ is the pseudogap temperature. For the hole doped cuprates
this field is of the order $\sim100$ T. In contrast, for the
electron-doped cuprates this field is of the order $\sim30$ T (and $T^*
\sim 30-40$ K), and so this is much more experimentally
accessible.\cite{kawakami} The field and temperature dependence of the
interlayer resistance for several superconducting organic charge
transfer salts\cite{zuo} is qualitatively similar to that for the
cuprates. In particular, for temperatures less than the zero-field
transition temperature and fields larger than the upper critical field,
negative magnetoresistance is observed for fields perpendicular to the
layers. A possible explanation is that, as in the cuprates, there is a
suppression of the density of states near the Fermi energy, and the
associated pseudogap decreases with increasing magnetic field.

A Nernst experiment can be used to probe whether there are
superconducting fluctuations in the pseudogap phase, as has been done
in the cuprates.\cite{wang} This experiment is particularly important
in understanding the relation between the pseudogap and
superconductivity.

One could also study the pressure dependence of the linear coefficient
of heat capacity $\gamma$. Since $\gamma$ is proportional to the
density of states at the Fermi energy, a detailed mapping of
$\gamma(P)$ would be an important probe for the study the pseudogap.
Finally, measurements of the Hall effect have also led to important
insights into the pseudogap of the cuprates\cite{timusk} therefore
perhaps the time is ripe to revisit these experiments in the organic
charge transfer salts.


\begin{acknowledgements}
The authors acknowledge stimulating discussions with Arzhang
Ardavan, Ujjual Divakar, John Fj\ae restad, David Graf, Anthony
Jacko, Moon-Sun Nam, Rajiv Singh, and Pawel Wzietek. We are grateful
to Ujjual Divakar and David Graf for critically reading the
manuscript. This work was funded by the Australian Research Council.
\end{acknowledgements}

\appendix

\section{Vertex corrections and the dynamic spin susceptibility for
Strongly Correlated Electrons}\label{appendix:vertex}

We consider a strongly interacting electron system and derive the real
and imaginary parts of the dynamic susceptibility. We show that under
some quite general (but specific) conditions that the Korringa ratio is
unity. Many definitions are simply stated in this appendix since most
are derived more fully in any number of textbooks (for example Ref.
\onlinecite{Mahan}). The general expression for the dynamic
susceptibility in Matsubara formalism is given by
\begin{equation}
\chi_{\alpha \beta}({\bf q},i\omega_n) = \int_0^\beta{d\tau e^{i
\omega_n \tau}\langle T_\tau m_\alpha({\bf q},\tau) m_\beta(-{\bf
q},0)\rangle}, \label{chi}
\end{equation}
where $\beta=1/k_B T$ is the inverse temperature, $\tau$ is the
imaginary time, $\omega_n=(2n+1)\pi k_B T/\hbar$ is the Matsubara
frequency, $m_\alpha$ is the component of magnetization in the $\alpha$
direction, and $T_\tau$ is the (imaginary) time ordering operator. In
order to consider $\chi_{-+}({\bf q},\omega)$ we define the operators:
\begin{eqnarray}
m_-({\bf q},\tau) &=& \frac{\hbar\gamma_e}{\sqrt{2}}\sum_{{\bf p}}
c_{{\bf p}+{\bf q},\downarrow}^\dagger(\tau) c_{{\bf
p},\uparrow}(\tau), \label{m} \\
m_+({\bf q},\tau) &=& \frac{\hbar\gamma_e}{\sqrt{2}}\sum_{{\bf p}}
c_{{\bf p}+{\bf q},\uparrow}^\dagger(\tau) c_{{\bf
p},\downarrow}(\tau). \label{mm}
\end{eqnarray}
Upon substituting (\ref{m}) and  (\ref{mm}) into (\ref{chi}) and
performing the appropriate Wick contractions on the operators one finds
that
\begin{eqnarray}
\chi_{-+}({\bf q},i\omega_n) =
\frac{\hbar^2\gamma_e^2}{2}\int_0^\beta{d\tau e^{i \omega_n
\tau}\Gamma({\bf p}+{\bf q},-\tau;{\bf p},\tau)}\nonumber\\\times
G({\bf p}+{\bf q},-\tau)G({\bf p},\tau) \label{chi_zz}
\end{eqnarray}
where $\Gamma({\bf q},\tau;{\bf p},\tau')$ is the vertex function,
$G({\bf p},ip_n)$ is the full interacting Green's function given by
\begin{equation}
G({\bf p},\tau) = \frac{G_0({\bf p},\tau)}{1-G_0({\bf
p},\tau)\Sigma({\bf p},\tau)},
\end{equation}
$G^0({\bf p},\tau)$ is the non interacting Green's function, and
$\Sigma({\bf p},\tau)$ is the self energy. The $\tau$ integration
can be evaluated by first transforming the integrand in Eq.
(\ref{chi_zz}) into momentum space. This gives
\begin{eqnarray}
\chi_{-+}({\bf q},i\omega_n) &=&
\frac{\hbar^2\gamma_e^2}{2\beta}\sum_{{\bf p},ip_m}\Gamma({\bf p}+{\bf
q},ip_m;{\bf p},ip_m+i\omega_n)\nonumber\\&&\times G({\bf p}+{\bf
q},ip_m)G({\bf p},ip_m+i\omega_n)\label{chi_zz_2}
\end{eqnarray}
where $\Gamma({\bf q},i\omega_n;{\bf p},i\omega_n')$ is the Fourier
transform of $\Gamma({\bf q},\tau;{\bf p},\tau')$ and $G({\bf
p},ip_n)$ given by
\begin{equation}
G({\bf p},ip_n)=\frac{1}{ip_n-\varepsilon_{\bf p}-\Sigma({\bf
p},ip_n)},
\end{equation}
where $\varepsilon_{\bf p}$ is the dispersion of the non-interacting
system. To evaluate the Matsubara summation, it is convenient to
express the full interacting Green's function using the spectral
representation
\begin{equation}
G({\bf p},ip_n) =
\int_{-\infty}^{\infty}{\frac{dE_1}{2\pi}\frac{A_s({\bf
p},E_1)}{ip_n-E_1}}, \label{lehmann}
\end{equation}
where $A_s({\bf p},E_1)$ is the spectral function given by
\begin{equation}
A_s({\bf p},E) = \frac{-2\mathrm{Im}\Sigma({\bf p},E)}
{(E-\varepsilon_{\bf p}-\mathrm{Re}\Sigma({\bf
p},E))^2+(\mathrm{Im}\Sigma({\bf p},E))^2}. \label{spectral}
\end{equation}
Substituting (\ref{lehmann}) into (\ref{chi_zz_2}), the dynamic
susceptibility becomes
\begin{eqnarray}
\chi_{-+}({\bf q},i\omega_n) &=&
\frac{\hbar^2\gamma_e^2}{2\beta}\sum_{{\bf p},m}\int_{-\infty}^{\infty}
\frac{dE_1}{2\pi}\frac{dE_2}{2\pi}\nonumber\\
&&\times\nonumber\Gamma({\bf p}+{\bf q},ip_m;{\bf p},ip_m+i\omega_n)
\\&&\times\frac{A_s({\bf p}+{\bf q},E_1)A_s({\bf
p},E_2)}{(ip_m-E_1)(ip_m+i\omega_n-E_2)}.\nonumber\\
\end{eqnarray}
At this stage we neglect vertex corrections, that is we set
$\Gamma({\bf p}+{\bf q},ip_n;{\bf p},ip_n+i\omega_n)=1$ for all ${\bf
p}$, ${\bf q}$, $p_n$, and $\omega_n$. After performing the Matsubara
sum
and analytical continuation $i\omega_n \to \omega + i\eta$, the dynamic
susceptibility is given by
\begin{eqnarray}
\chi_{-+}({\bf q},\omega) &=& \frac{\hbar^2\gamma_e^2}{2} \sum_{{\bf
p}}\int_{-\infty}^{\infty}\frac{dE_1}{2\pi}\frac{dE_2}{2\pi}A_s({\bf
p}+{\bf q},E_1)
\nonumber\\
&& \times A_s({\bf p},E_2)
\frac{n_F(E_1)-n_F(E_2)}{\hbar\omega+E_1-E_2+i\eta},\label{chi_final}
\end{eqnarray}
where $n_F(E)$ is the Fermi function.

First we discuss the imaginary part of $\chi_{-+}({\bf q},\omega)$.
Using the well known relation $1/(x+i\eta) = P(1/x) - i \pi
\delta(x)$, where $P(y)$ denotes the principal value, the imaginary
part of $\chi_{-+}({\bf q},i\omega_n)$ in the limit of small
frequency $\omega$ is given by
\begin{eqnarray}
&&\lim_{\omega \to 0} \frac{\chi''_{-+}({\bf q},\omega)}{\omega}\nonumber\\
&&= \frac{\hbar^2\gamma_e^2}{2}\sum_{{\bf p}}
\int_{-\infty}^{\infty}\frac{dE}{4\pi}A_s({\bf p}+{\bf q},E)A_s({\bf
p},E) \left(-\frac{\partial n_F}{\partial
E}\right).\nonumber\\\label{im_chi_final}
\end{eqnarray}
The nuclear spin relaxation rate is obtained by summing the dynamic
susceptibility over all q  [c.f., Eq. (\ref{t1t})] thus we find that
\begin{eqnarray}
\frac1{T_1T} &=& \frac{k_B |A|^2}{\hbar}
\int_{-\infty}^{\infty}\frac{dE}{4\pi}\sum_{{\bf p}}\sum_{{\bf q}}A_s({\bf p},E)A_s({\bf p}+{\bf q},E)\nonumber\\
&&\times\left(-\frac{\partial n_F}{\partial E}\right),
\end{eqnarray}
where we have assumed a contact, i.e. momentum independent, hyperfine
coupling. This expression can be written in the more intuitive form
\begin{eqnarray}
\frac1{T_1T} = \frac{k_B |A|^2}{\hbar}
\int_{-\infty}^{\infty}\frac{dE}{4\pi}\tilde{\rho}^2(E)
\left(-\frac{\partial n_F}{\partial E}\right),\label{t1t_local1}
\end{eqnarray}
where
\begin{equation}
\tilde{\rho}(E)=\sum_{\bf p} A_s({\bf p},E) \label{dos}
\end{equation}
is the full interacting density of states. At temperatures small enough
so that $\tilde{\rho}(E)$ varies little with energy within $k_B T$ of
the Fermi energy, $E_F$, the expression can be further simplified to
\begin{equation}
\frac1{T_1T} \simeq \frac{k_B |A|^2
\tilde{\rho}^2(E_F)}{4\pi\hbar}.\label{t1t_local2}
\end{equation}

Within the approximation of neglecting vertex corrections the real
part of the dynamic susceptibility is given by
\begin{eqnarray}
\chi'({\bf q},\omega) &=& \frac{\hbar^2\gamma_e^2}{2}\sum_{{\bf
p}}\int_{-\infty}^{\infty}
\frac{dE_1}{2\pi} A_s({\bf p}+{\bf q},E_1)\nonumber\\
&&\times \int_{-\infty}^{\infty}\frac{dE_2}{2\pi}F({\bf
p},E_1,E_2,\omega,T) \label{real_chi_final}
\end{eqnarray}
with
\begin{equation}
F({\bf p},E_1,E_2,\omega,T) = \frac{A_s({\bf
p},E_2)[n_F(E_1)-n_F(E_2)]}{\hbar\omega+E_1-E_2}.
\end{equation}
To perform the integration over $E_2$, we first make the change of
variable $x=\omega+E_1-E_2$ and take the limit $\omega \to 0$. Thus
the integral over $E_2$ becomes
\begin{eqnarray}
&&\int_{-\infty}^{\infty} \frac{dx}{2\pi}
F({\bf p},E_1,x,\omega,T)\nonumber\\
&&= -\int_{-\infty}^{\infty}\frac{dx}{2\pi}A_s({\bf p},E_1-x)
\frac{n_F(E_1)-n_F(E_1-x)}{x}\nonumber\\
&&=\left(-\frac{dn_F}{dE_1}\right)
\int_{-\infty}^{\infty}\frac{dx}{2\pi}A_s({\bf p},E_1-x).
\end{eqnarray}
By using sum rule
\begin{equation}
\int_{-\infty}^{\infty} \frac{dy}{2\pi} A_s({\bf p},y) = 1,
\end{equation}
$\chi'_{-+}({\bf q},0)$ then becomes
\begin{eqnarray}
\chi'_{-+}({\bf
q},0)=\frac{\hbar^2\gamma_e^2}{2}\int_{-\infty}^{\infty}\frac{dE}{2\pi}
\sum_{{\bf p}}A_s({\bf p}+{\bf q},E)\left(-\frac{\partial
n_F}{\partial E}\right)\nonumber\\.
\end{eqnarray}
The Knight shift is obtained by setting ${\bf q}=\bf0$ [c.f., Eq.
(\ref{ks})] and is
\begin{equation}
K_s = \frac{|A|\gamma_e}{2\gamma_N }
\int_{-\infty}^{\infty}\frac{dE}{2\pi}
\tilde{\rho}(E)\left(-\frac{\partial n_F}{\partial
E}\right).\label{ks_local1}
\end{equation}
At temperatures sufficiently low that $\tilde{\rho}(E)$ varies little
with energy within $k_B T$ near the Fermi energy, the Knight shift is
given by
\begin{equation}
K_s \simeq
\frac{|A|\gamma_e\tilde{\rho}(E_F)}{4\pi\gamma_N}.\label{ks_local2}
\end{equation}
Using Eqs. (\ref{t1t_local2}) and (\ref{ks_local2}), the Korringa ratio
for interacting electrons with a contact hyperfine coupling and
neglecting vertex corrections is
\begin{eqnarray}
\mathcal{K} &=& \frac{\hbar}{4\pi
k_B}\left(\frac{\gamma_e}{\gamma_N}\right)^2 \frac{1}{T_1T
K_s^2}\nonumber\\
&=& \frac{\hbar}{4\pi k_B}\left(\frac{\gamma_e}{\gamma_N}\right)^2
\frac{k_B |A|^2\tilde{\rho}^2(E_F)}{4\pi\hbar}
\left(\frac{4\pi\gamma_N}{|A|\gamma_e\tilde{\rho}(E_F)}\right)^2\nonumber\\
&=& 1.\label{korringa_local}
\end{eqnarray}

For non interacting electrons, the self energy $\Sigma({\bf
q},\omega)=0$;  expressions Eqs. (\ref{t1t_local1}) and
(\ref{ks_local1}) are still valid and one only need replace
$\tilde{\rho}(E)$ with the non interacting density of states $\rho(E)$.
On the temperature scale over which the density of states varies little
with energy, the Korringa ratio for free electron gas ${\cal
K}_\mathrm{free}=1$. By comparing Eqs. (\ref{korringa_local}) and
${\cal K}_\mathrm{free}=1$ we see that any deviation of the Korringa
ratio from one must be caused by either vertex corrections or the
q-dependence of the hyperfine coupling.

\section{Estimation of effect of thermal expansion of the lattice on the Knight
shift}\label{sect:estimation}

In this appendix we describe how we obtained our estimates of the
isothermal compressibility the linear coefficient of thermal expansion
and the pressure dependence of the Knight shift. By feeding these
estimates into Eq. (\ref{correction}) we estimate the correction to the
Knight shift required because measurements are generally performed at
constant pressure whereas calculations are most naturally performed at
constant volume. The difference between these two versions of the
Knight shift can be quite significant as can be seen in Fig.
\ref{fig:correction}.

First, let us discuss the first term in the integrand in Eq.
(\ref{correction}). We observed that $K_s^p$ can be rewritten as $K_s^p
= [\hbar \gamma_e^2/(4\pi k_B \gamma_N^2 T_1T{\cal K})]^{1/2}$. Using
Mayaffre's data,\cite{mayaffre} we extracted the pressure dependence of
$1/T_1T$ at constant temperature and estimated that $(1/T_1T)^{1/2}$ is
roughly linear with pressure: $(1/T_1T)^{1/2} \sim -3$x$10^{-5} P$ for
$T_1$ in second, $T$ in Kelvin, and $P$ in bar. We then used this
result and the Korringa value for non interacting electron gas to
calculate $\left(\partial K_s^p/\partial P\right)_T$ for \kbr~which is
found to be around $-3$x$10^{-8}$/bar. One can compare the value
obtained here with $\left(\partial K_s^p/\partial P\right)_T$ obtained
from the pressure dependence study on effective mass in
\kncs~compound\cite{caulfield} which is presumably more reliable. Since
$K_s$ should be proportional to the effective mass, the quantity of
interest $\left(\partial K_s^p/\partial P\right)_T$ can be estimated
from the pressure dependence of the effective mass data which yields a
value around $-6$x$10^{-8}$/bar. The two estimates agree to with a
factor of two.

Next we need to obtain a value for lattice isothermal
compressibility. We use the analytical expression obtained from
DMFT\cite{hassan}
\begin{equation}
(\mathscr{K}v)^{-1} = \frac{B_0}{v_0} - \left(\frac{\nu
D_0}{v_0}\right)^2\chi_\mathrm{el},
\end{equation}
where $\mathscr{K}$ is the inverse isothermal compressibility $(v
\partial P/\partial v)^{-1}$, $v_0$ is the reference unit-cell volume,
$v$ is the unit-cell volume under pressure, $B_0$ is a reference bulk
elastic modulus, $D_0$ is a reference bandwidth, $\nu$ is a parameter
that characterizes the change in the bandwidth under pressure, and
$\chi_\mathrm{el}$ is the electronic susceptibility. We estimated that
for \kcl, $v_0=1700~\AA^3$, $B_0=122$ kbar, $D_0=0.13$ eV, and $D_0
\chi_\mathrm{el} \sim 1$. Putting everything together, the order of
magnitude of the lattice isothermal compressibility for \kcl~is around
$-10^5$ bar. Although we are not aware of any measurements of the
isothermal compressibility of \kcl~ systematic axial pressure
studies\cite{kondo} on $\alpha$-(BEDT-TTF)NH$_4$Hg(NCS)$_4$ found that
the isothermal compressibility is of order $-10^5$ bar, a value which
is very close to our crude estimate for \kcl.

The temperature dependence of the thermal expansion at constant
pressure has been measured by M\"{u}ller {\it et
al}.\cite{muller:lattice} They found that \kcl, \kncs, and undeuterated
and fully deuterated \kbr~all have a relatively temperature independent
thermal expansion above about 80~K while complicated features are
observed below 80 K associated with glassy transitions and the
many-body behavior of these materials. Since we are only interested in
getting an order of magnitude, we neglect the complicated temperature
dependence observed in \cation-X and approximate the thermal expansion
as a constant. The value extracted from M\"uller {\it et al}.'s data is
around $10^{-4}$~K$^{-1}$.

\section{Estimation of Errors in Figure
\ref{fig:phase}}\label{sect:errors}

In this appendix we outline the procedure use to estimate the errors on
the data presented in Fig \ref{fig:phase}. Let us consider a given set
of data, for example $T_\c (P)$. To a reasonable degree of accuracy,
$T_\c$ for \kpx~ decreases linearly with increasing pressure, i.e.
$T_\c = a P + b$ where $a$ and $b$ are the coefficients obtained from
fitting the expression to the data. In a typical pressure measurement,
there will be some uncertainties in the pressure ($\Delta P$) which may
be caused by systematic errors due to the uncertainties in the pressure
calibration. The size of $\Delta P$ will depend on a specific apparatus
used in the experiment. For example, a helium gas pressure system would
have uncertainties as large as 0.1 kbar while a clamped pressure cell
which uses oil pressure medium may have uncertainties as large as 1
kbar.\cite{murata,graf} Knowing $\Delta P$, we can estimate the
uncertainty in $T_\c$ when it is compared with another data set from a
different experiment taken by a different group, say $T_\res (P)$. This
is done by calculating the following:
\begin{equation}
\Delta T_\c = \frac{d T_\c(P)}{dP} \Delta P.\label{delta}
\end{equation}
Another way to estimate $\Delta T_\c$ is to calculate $d T_\c(P)/dP$
from the discontinuity in the thermal expansion and specific heat by
using the Ehrenfest relation.\cite{muller:sm} These two methods give
similar results. For \kbr, $d T_\c(P)/dP$ is estimated to be around
$-2.4$ K/kbar from the first method while it is found to be around
$-2.2$ K/kbar from the Ehrenfest relation. This procedure is repeated
for other data sets, $T_\res (P)$, $T_\us (P)$, and $T_\nmr (P)$
whenever applicable which leads to $T_\c \pm \Delta T_\c$, $T_\res \pm
\Delta T_\res$, $T_\us \pm \Delta T_\us$, and $T_\nmr \pm \Delta
T_\nmr$. We then tabulate $T_\res, T_\us,$ and $T_\nmr$ with respect to
$T_\c$ for a given pressure. The results are shown in Fig
\ref{fig:phase}. In some cases we were not able to obtain a reasonable
fit because either the data are very scattered or there are not enough
data to perform fit. This is the reason for the absence of error bars
on the vertical axis for some data points in Fig \ref{fig:phase}.

\bibliographystyle{apsrev}
\bibliography{reference}

\end{document}